\documentclass[preprint2]{aastex631}

\shorttitle{UBAD}
\shortauthors{Munn et al.}

\graphicspath{{./}{figures/}}

\epsscale{1.176}

\begin{document}

\title{Accurate Ground-based Astrometry of Naked-eye Stars:  The United States Naval
  Observatory Bright-Star Astrometric Database}

\author[0000-0002-4603-4834]{Jeffrey A. Munn}
\affiliation{U.\ S.\ Naval Observatory, Flagstaff Station,
  10391 W.\ Naval Observatory Road, Flagstaff, AZ 86005-8521, USA}
\author[0000-0001-5912-6191]{John P. Subasavage}
\affiliation{The Aerospace Corporation, 2310 E.\ El Segundo Boulevard,
  El Segundo, CA 90245, USA}
\affiliation{U.\ S.\ Naval Observatory, Flagstaff Station,
  10391 W.\ Naval Observatory Road, Flagstaff, AZ 86005-8521, USA}
\author{Hugh C. Harris}
\affiliation{U.\ S.\ Naval Observatory, Flagstaff Station,
  10391 W.\ Naval Observatory Road, Flagstaff, AZ 86005-8521, USA}
\author{Trudy M. Tilleman}
\affiliation{U.\ S.\ Naval Observatory, Flagstaff Station,
  10391 W.\ Naval Observatory Road, Flagstaff, AZ 86005-8521, USA}
\correspondingauthor{Jeffrey A. Munn}
\email{jeffrey.a.munn2.civ@mail.mil}

\begin{abstract}
We present the
United States Naval Observatory (USNO) Bright-Star Astrometric Database (UBAD), a current-epoch high-accuracy astrometric catalog.  The catalog consists of 364 bright northern hemisphere stars, including all but five such stars with either $V < 3.5$ or with $I < 3.2$ and $V < 6$, as well as a large fraction of slightly fainter stars;
36 of the brightest catalog stars are not included in Gaia Early Data Release 3 (EDR3). 
Observations were conducted with the USNO, Flagstaff Station, Kaj Strand 61-inch Astrometric Reflector. Target stars were imaged through a small 12.5-magnitude neutral-density spot, while the remainder of the stars in the field of view were unattenuated.  This allowed for unsaturated images of the bright target stars to be calibrated directly against much fainter reference stars from Gaia EDR3.  The median position errors are
1.9~mas in both right ascension and declination at the catalog epoch of 2017.0, with 90\% of catalog stars
having errors less than 2.6 mas; systematic errors are
1 -- 3 mas.  Combining UBAD observations with Hipparcos-2 positions yields proper motions with median errors of 0.045 and
0.049~mas~year$^{-1}$ in right ascension and declination, respectively, with 90\%
of stars having errors less than 0.1~mas~year$^{-1}$; systematic errors are about 0.1~mas~year$^{-1}$.  Single-frame accuracy for
positions of the target stars is typically 5 -- 6 mas.
Gaia EDR3 astrometry for these bright stars, which are heavily saturated in the Gaia
observations, is validated over the magnitude range $2 \lesssim G \lesssim 6$.
\end{abstract}

\section{Introduction}

The brightest stars in the sky, including those visible to the naked eye,
are used as reference sources in many government and commercial applications,
including star trackers, navigation systems, and weapon systems.  Before
the European Space Agency's (ESA) Gaia mission \citep{2016A&A...595A...1G}, the most accurate astrometry
available for these bright sources was the Hipparcos-2 catalog \citep{2007ASSL..350.....V,2007A&A...474..653V}, another product of ESA.  For stars brighter than $H_p = 6$, Hipparcos-2 has typical errors for well-behaved stars of 0.1 -- 0.2~mas in position and 0.1 -- 0.3~mas~year$^{-1}$ in
proper motion at the catalog epoch of 1991.25.
Propagating the Hipparcos-2 positions over 30 years to today, the proper motion
errors lead to position errors at the current epoch of as great as 9 mas, though considerably worse for stars that do not have well-behaved solutions in Hipparcos-2, and the errors
increase linearly going forward in time.  Further, the Hipparcos satellite took observations for only 3.5 years \citep{1997A&A...323L..49P} and thus was not sensitive to detecting multiple systems with periods appreciably longer than its mission lifetime.

During the building stage of the Gaia mission, the bright magnitude limit was defined to be $G \sim 5.7$.  Given the importance of accurate bright-star astrometry, in particular for
United States Department of Defense applications, the United States Naval Observatory (USNO), Flagstaff Station, commenced a program in 2012 to obtain accurate current-epoch ground-based astrometry of bright northern hemisphere stars, those that Gaia would not observe, including all such stars in the magnitude range $3 < V < 6$.  The basic technique was to observe the bright target star through a small neutral-density spot located in front of the detector, such that only the light from the target star and its immediate vicinity is attenuated, while the rest of the field of view remains unattenuated.  The first observations were conducted with a 9-magnitude neutral-density spot, allowing the target star to be directly imaged with, and thus calibrated against, reference stars more than nine magnitudes fainter.  These reference stars would be unsaturated on the Gaia instrument and thus would eventually have accurate Gaia astrometry once Gaia released its first data.  This technique, using the same 9-magnitude neutral-density spot in a different camera, was first
developed at USNO to validate Hipparcos astrometry \citep{1997ESASP.402..105H} and later
to validate a program of wide-angle absolute astrometry with the Navy Precision Optical
Interferometer \citep{ADA511248,2011AAS...21734804Z}.

In 2014 the Gaia Collaboration announced that they had modified the object detection and collection algorithms on the spacecraft to allow the collection of data for the brightest stars, though it was unclear whether they would be able to achieve the same astrometric accuracy for these stars given that their images would be heavily saturated \citep{2014SPIE.9143E..0YM,2016SPIE.9904E..2ES}.  In light of these developments, we decided to change the focus of our work to the very brightest stars, those that would be most challenging for Gaia.  It was also realized that while Gaia was likely to provide more accurate astrometry on even these bright stars than we could achieve from the ground, our survey could provide one of the few external validations of Gaia astrometry for bright stars.
The 9-magnitude neutral-density spot was replaced with a 12.5-magnitude spot and a much brighter sample of stars was targeted.

This paper presents the results of the latter survey and compares it with Gaia Early Data Release 3 \citep[EDR3,][]{2021A&A...649A...1G,2021A&A...649A...2L,2021A&A...649A..11R,2021A&A...649A...5F}.  Section~\ref{sec:observations} describes the instrumentation and observations.  Section~\ref{sec:processing} describes the processing of the individual observations and characterizes the quality of those observations.  Section~\ref{sec:catalog-astrometry} describes the production of the catalog from the individual observations and compares it with Gaia and Hipparcos-2.  Section~\ref{sec:catalog} presents the catalog, and Section~\ref{sec:conclusion} summarizes our results.

\section{Observations}
\label{sec:observations}

The USNO Bright-Star Astrometric Database (UBAD) targets all northern hemisphere ($\delta > 0\arcdeg$) stars in the original Hipparcos catalog with $V < 3.5$, or with $I < 3.5$ and $V < 6$, with the following exceptions:
\begin{enumerate}
    \item Polaris, at a declination $\delta = 89\arcdeg 16\arcmin$, lies above the useful declination limit of the telescope used for the survey, and thus was excluded;
    \item 46 stars that were previously observed in the first survey with the 9-magnitude neutral-density spot were excluded --- all have $V > 3.5$, and only four have $I < 3.2$;
    \item an additional five stars with $6 < V < 7$ were included, to extend the magnitude range of overlap with Gaia.
\end{enumerate}
The final sample numbers 364 stars.

Observations were conducted with the USNO, Flagstaff Station, Kaj Strand 61-inch Astrometric Reflector \citep{1964S&T....27..204S}.  The telescope, commissioned in 1963, was designed specifically for precise narrow-field differential astrometry and has a long history determining stellar parallaxes in both the optical \citep[][and references therein]{2017AJ....154..147D} and infrared \citep{2004AJ....127.2948V}.  The camera was built in-house using a 2048 $\times$ 4102 e2v CCD with 15-micron pixels, which in the 61-inch focal plane yields a pixel scale of 0.2025 arcsec pixel$^{-1}$.  A clear filter covers the detector with a 5-mm diameter 12.5-magnitude neutral-density spot deposited in the center on the sky-facing surface of the filter.  The target star is observed through the spot, which has measured attenuations of 12.48 magnitudes in $V$ and 12.89 magnitudes in $I$.

A total of 4807 observations were obtained between March 2016 and October 2020.
All observations were taken using either an SDSS $i$ or $z$
filter, mostly within an hour of the meridian, to minimize differential
chromatic refraction (DCR).  At a minimum, each star was targeted with two
visits on different nights, with two observations per visit.

After the initial set of data was processed and the final catalog generated, we decided to observe a set of fainter stars to further extend the magnitude range of overlap with Gaia and thus better constrain any dependence on magnitude in Gaia's bright-star astrometry. Thus, an additional 151 observations were obtained in $i$ between March and May 2021, targeting stars with $4.0 < G < 6.5$ (the survey was cut short due to a detector failure).  These observations are not used
in the UBAD catalog;  they are only used in Figure~\ref{fig:g-diff-I} below.

\section{Image Processing}
\label{sec:processing}
\subsection{Object Detection and Characterization}

\begin{figure}
\plotone{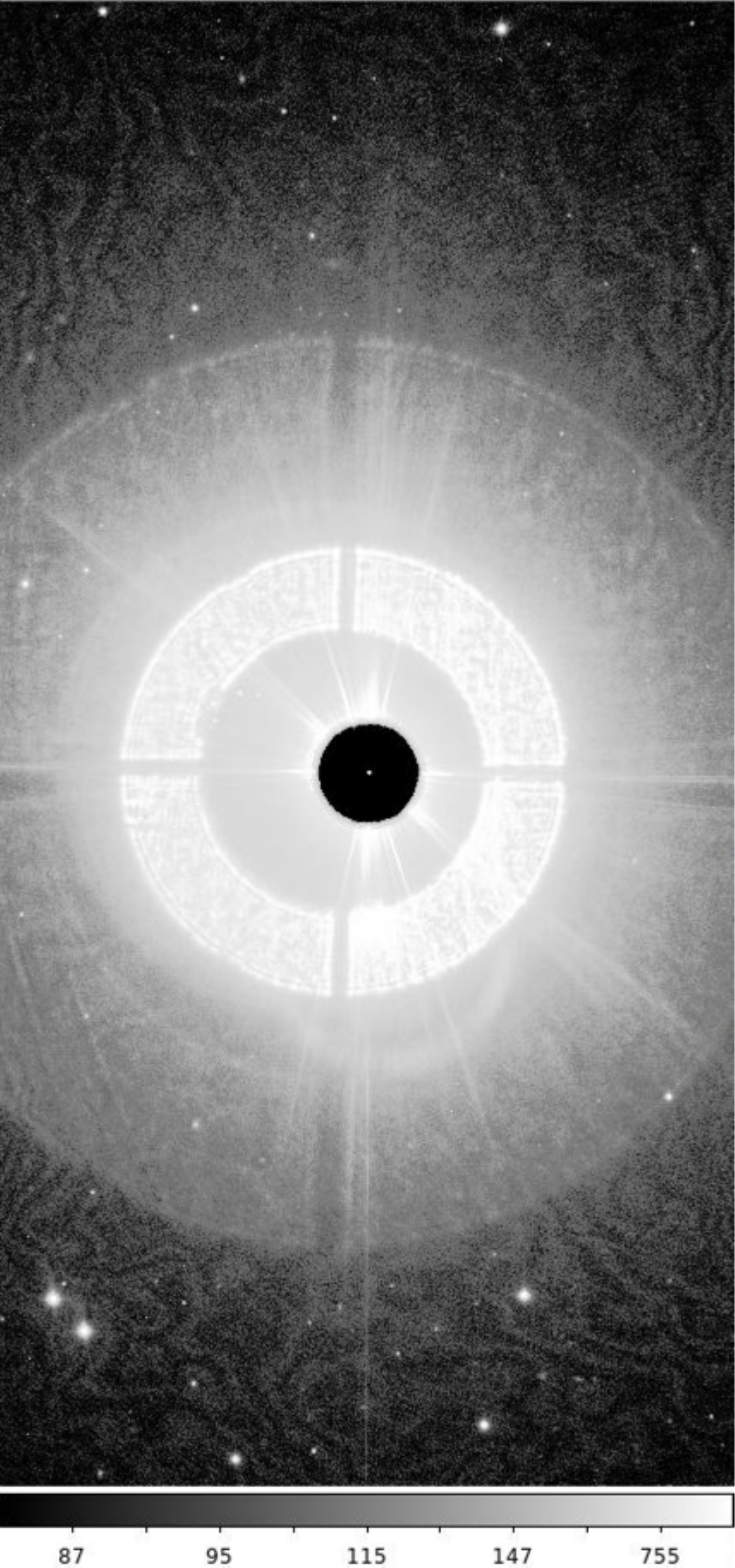}
\caption{Typical survey image, taken through the $i$ filter.  The color bar indicates the non-linear scaling used to highlight various features in the image.\label{fig:image}}
\end{figure}

\begin{figure}
\plotone{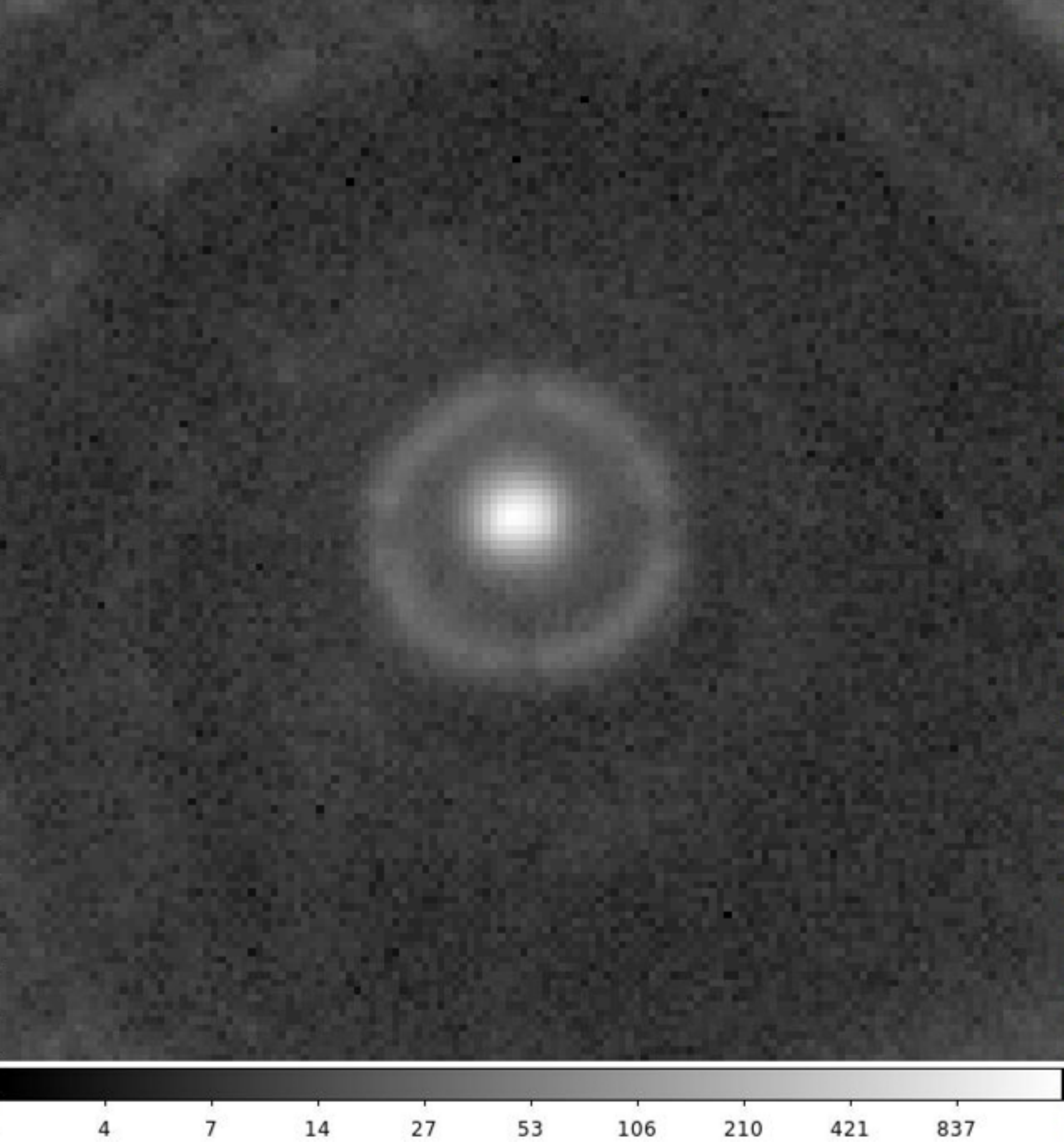}
\caption{Close-up under the neutral-density spot of the image displayed in
Figure~\ref{fig:image}, using a log scaling to highlight low-level features in the image.\label{fig:spot}}
\end{figure}

Figure~\ref{fig:image} displays a typical survey image, using a non-linear scaling to reveal various features.  The dark spot in the center of the frame is the area of the CCD covered
by the neutral-density filter.  The target bright star is visible in the center of the
spot, its flux attenuated by over 12 magnitudes.  Scattered light from this bright star is evident around the spot and
extending at a much lower level over about half the chip.  Two out-of-focus images of the target star from multiple reflections between surfaces in the camera dominate the central half of the chip.  Fringing is also evident.  A close-up of the portion of the image under the neutral-density
spot is shown in Figure~\ref{fig:spot}.  The ring around the target star is
present in all images, though its brightness is only a few percent of the peak
brightness.  The background is considerably less uniform than a typical astronomical
image.

The images are flat-fielded using a median of three dome flats.  As it is
impossible to get adequate counts under the neutral-density spot without
saturating the rest of the CCD, an archival flat produced before the spot was
installed is used for the region under the spot; there is no way to track
changing dust spots in that portion of the image.  SExtractor \citep{1996A&AS..117..393B} is used to detect
and measure stars on the images.  A smaller bin size for the sky is used than
usual, to better track the varying background.  SExtractor is run separately on the portion of the image
under the neutral-density spot, with tweaked deblending parameters so as not to
deblend the ring from the target star.  Centers are measured using
SExtractor's windowed first-order moments (XWIN\_IMAGE and YWIN\_IMAGE).

\subsection{Astrometric Calibration}
\begin{figure*}
\plotone{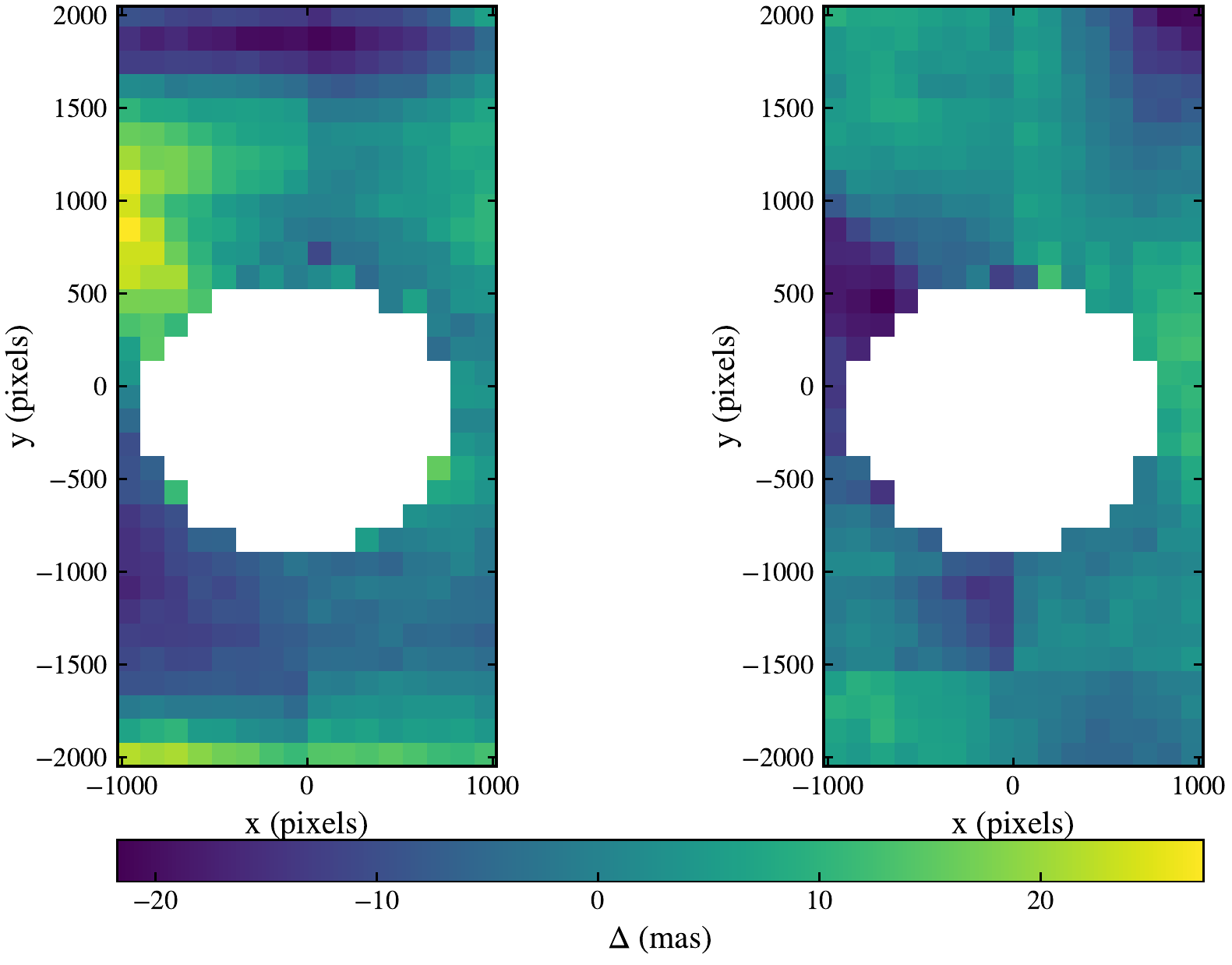}
\caption{Residual maps for observations taken in $i$ (catalog position minus observed position).  The left and right panels display the maps for right ascension and declination, respectively.  The CCD manufacturing process leads to discontinuities between regions on the CCD $1024 \times 512$ pixels in size, evident in the maps.  The large hole in the center of each map corresponds to the portion of the CCD not used in the astrometric calibrations due to the bright out-of-focus images of the target star.
\label{fig:rmap-I}}
\end{figure*}

\begin{figure*}
\plotone{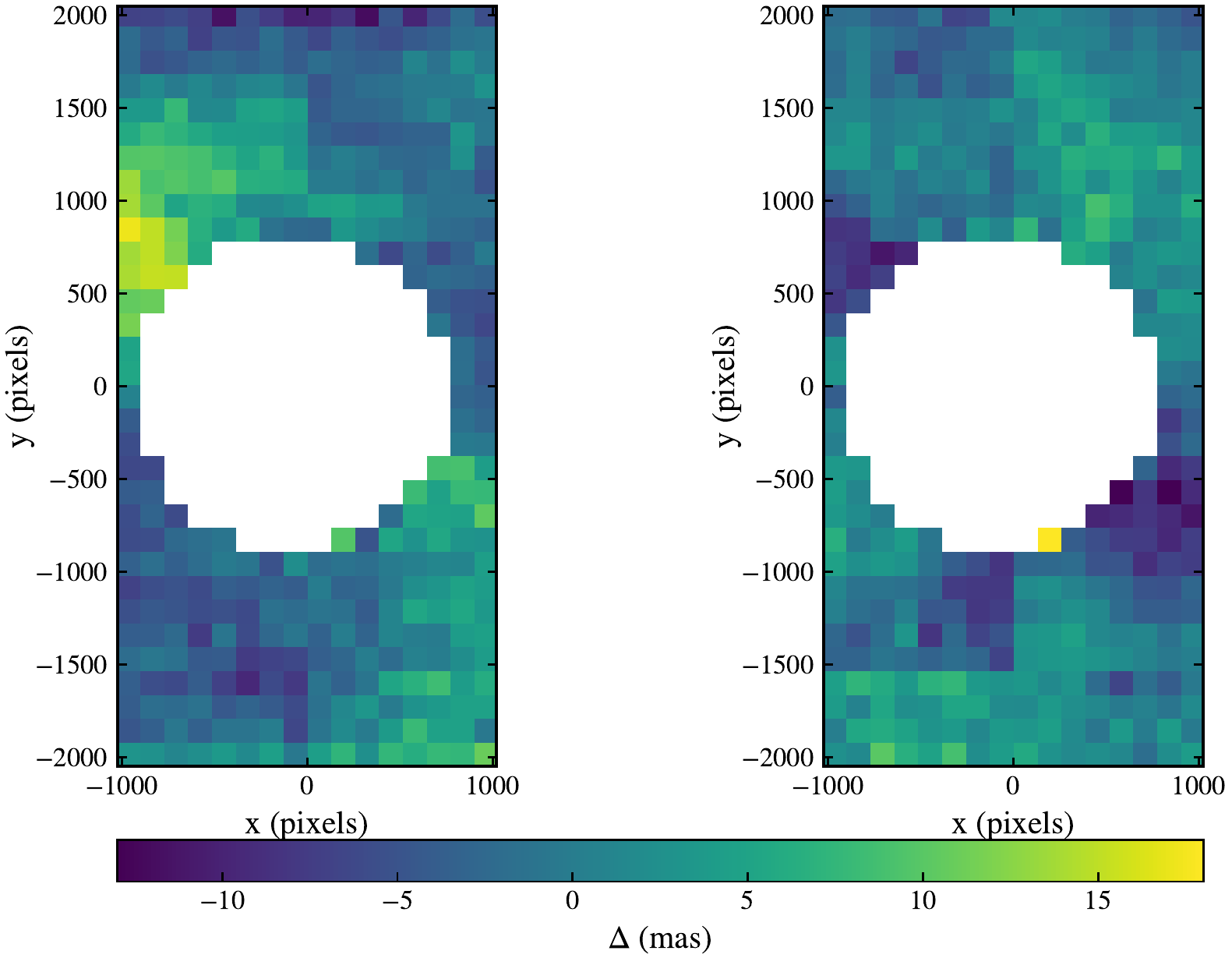}
\caption{Same as Figure~\ref{fig:rmap-I}, but for observations taken in $z$.\label{fig:rmap-Z}}
\end{figure*}

Each survey image was astrometrically calibrated against reference stars from
Gaia EDR3 using custom software, built using several invaluable community software packages highlighted below.
The following cuts were applied to the Gaia catalog to yield reliable
astrometric calibrators:
\begin{enumerate}
\item \textbf{astrometric\_params\_solved} = 31 or 95 (5- or 6-parameter
solution),
\item $\textbf{ruwe} < 1.4$,
\item $\textbf{ipd\_gof\_harmonic\_amplitude} < 0.15$,
\item $\textbf{ipd\_frac\_multi\_peak} <= 3$,
\item $\textbf{ipd\_frac\_odd\_win} < 10$.
\end{enumerate}
No stars were used within a large
circle on the images encompassing the brightest portion of the out-of-focus images of the target star.
The initial match of detected objects on the image against Gaia was performed using astroalign \citep{BEROIZ2020100384}, or in those cases where astroalign fails, with Astrometry.net \citep{2010AJ....139.1782L}.
A simple affine transformation was fit for each image;  no additional optics
terms were found necessary.
Fits were performed after first transforming Gaia coordinates to observed place, using place conversion routines provided by Astropy \citep{astropy:2013, astropy:2018}, including its interface to Essential Routines for Fundamental Astronomy \citep[PyERFA,][]{https://doi.org/10.5281/zenodo.4279480}, and corrected for polar motion using the Positional Astronomy Library \citep[PALpy,][]{2013ASPC..475..307J}.
Separate residual maps in $i$ and $z$ were created from the initial fits by calculating the clipped mean residual for the calibration stars, separately in right ascension and declination, binned in $128 \times 128$ pixel regions of the images.  The set of images used to generate the residual maps were restricted to those where 1) the rms of the residuals in both right ascension and declination were less than 12 mas, 2) the calibration used at least 30/25 stars for observations using the $i/z$ filter, and 3) the seeing was less than 1.5/1.7 arcsec for observations using the $i/z$ filter; looser constraints were used in $z$ due to the smaller number of observations taken in that filter. Figures~\ref{fig:rmap-I} and \ref{fig:rmap-Z} display the
residual maps in $i$ and $z$, respectively.  Residuals in $i$ have
peak-to-peak systematics of about -20 to 25 mas, while in $z$ they are
about -13 to 18 mas.
The $z$ residual map is considerably noisier as there were far
fewer observations in $z$ than $i$.
Separate residual maps in both bands were generated
spanning periods of several months, rather than over
the entire survey, to look for changes in the residual maps with time; no
significant changes were seen.  Thus,
a single residual map was used in each band for the entire survey.  The
pipeline was then rerun after correcting the on-chip positions, using corrections derived
by interpolating the residual maps with bivariate splines.

\begin{figure}
\plotone{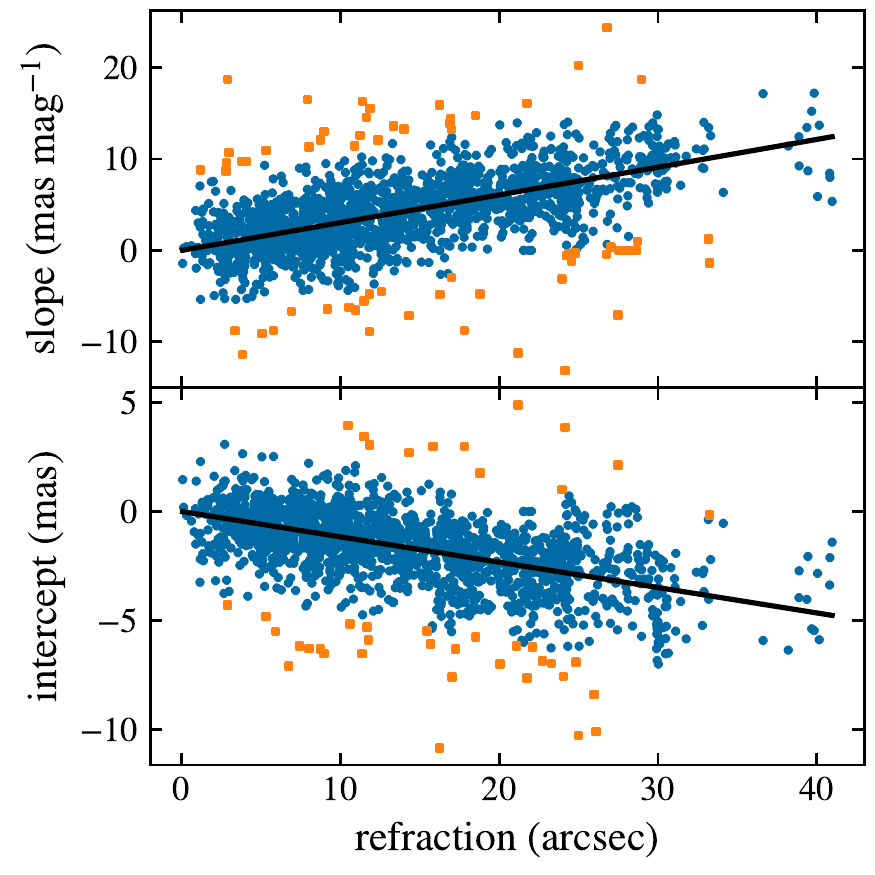}
\caption{The slope and intercept values for the DCR fits for the ensemble of
  $i$ images, plotted against refraction.  The black lines are the fits to
  the ensemble, used to apply DCR corrections in $i$.  Blue circles were
  included in the fits, orange squares were iteratively rejected as greater than
  3 $\sigma$ outliers.}.
\label{fig:dcr-I}
\end{figure}

\begin{figure}
\plotone{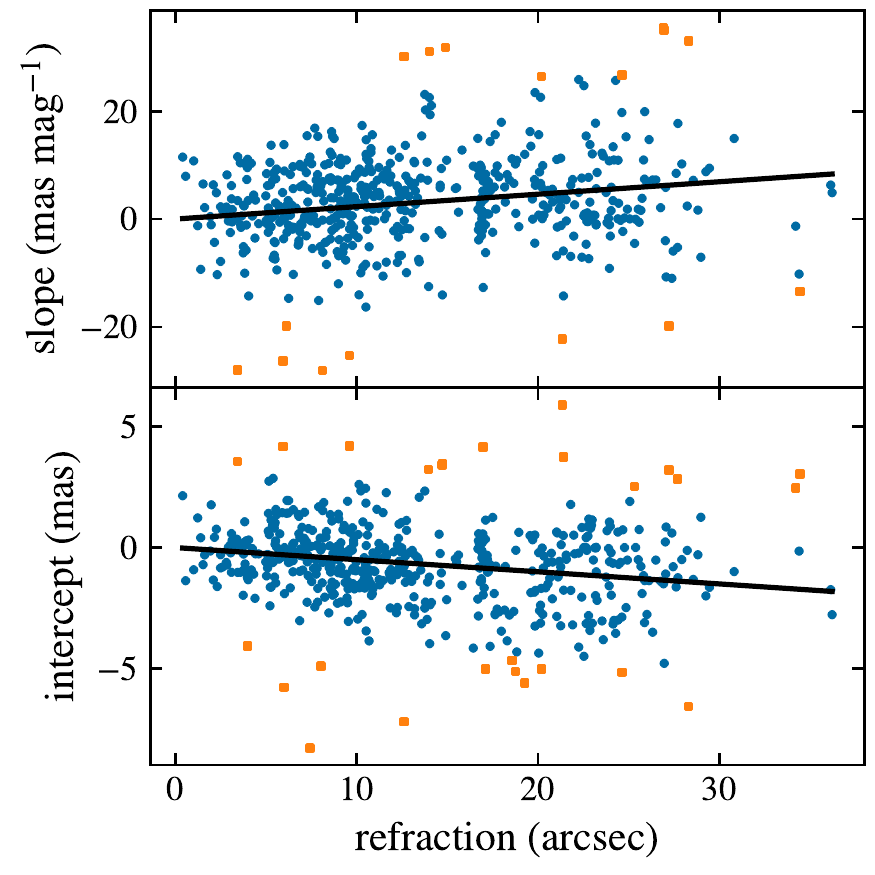}
\caption{Same as Figure~\ref{fig:dcr-I}, but for the $z$ images.}
\label{fig:dcr-Z}
\end{figure}

Next, corrections
for DCR were derived.  For each image, a line was fit to the residuals
along the parallactic angle versus the colors of the reference stars, using Pan-STARRS
\citep{2016arXiv161205560C,2020ApJS..251....3M} $r-i$ for images taken in $i$, and $i-z$ for images taken in $z$.  As this is a
noisy measurement for individual frames, the ensembles for the slope and
intercept values, limited to the same set of good images used to generate the residual maps, were then fit with a linear function against
refraction, separately in $i$ and $z$, with each fit forced to go through 0
at a refraction of 0.  The results are shown in Figures~\ref{fig:dcr-I} and
\ref{fig:dcr-Z}.  The fits yield the following corrections for DCR in $i$ and
$z$ as a function of refraction and star color:
\begin{eqnarray}
{\rm DCR}_i\ (\rm mas) & = & 0.3057\ \theta\ (r - i - 0.3893), \\
{\rm DCR}_z\ (\rm mas) & = & 0.2541\ \theta\ (i - z - 0.2220),
\end{eqnarray}
where $\theta$ is the refraction in arcseconds.

\begin{figure}
\plotone{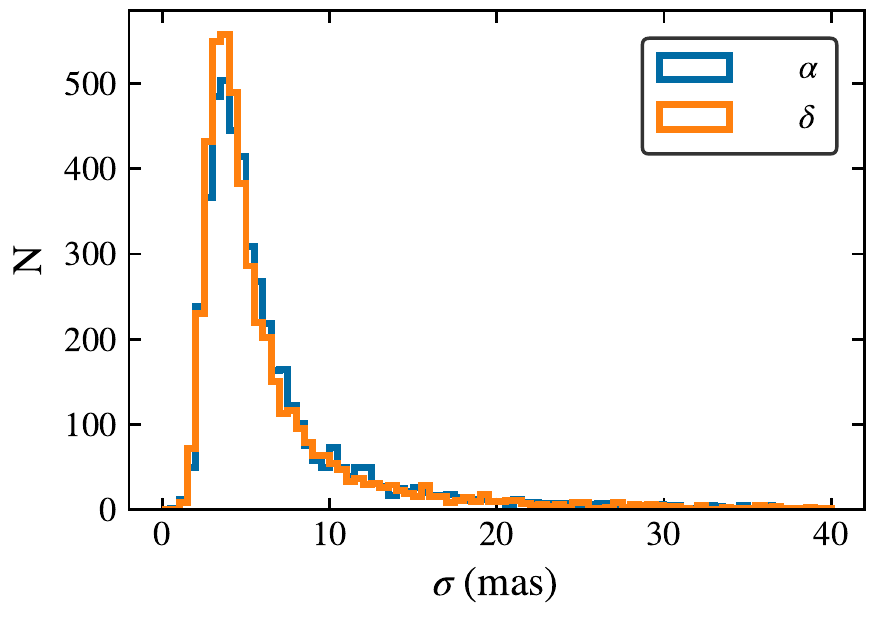}
\caption{Distribution of the rms residuals for the astrometric solutions for
all images in the survey, separately in right ascension and declination.}
\label{fig:rms}
\end{figure}

\begin{figure}
\plotone{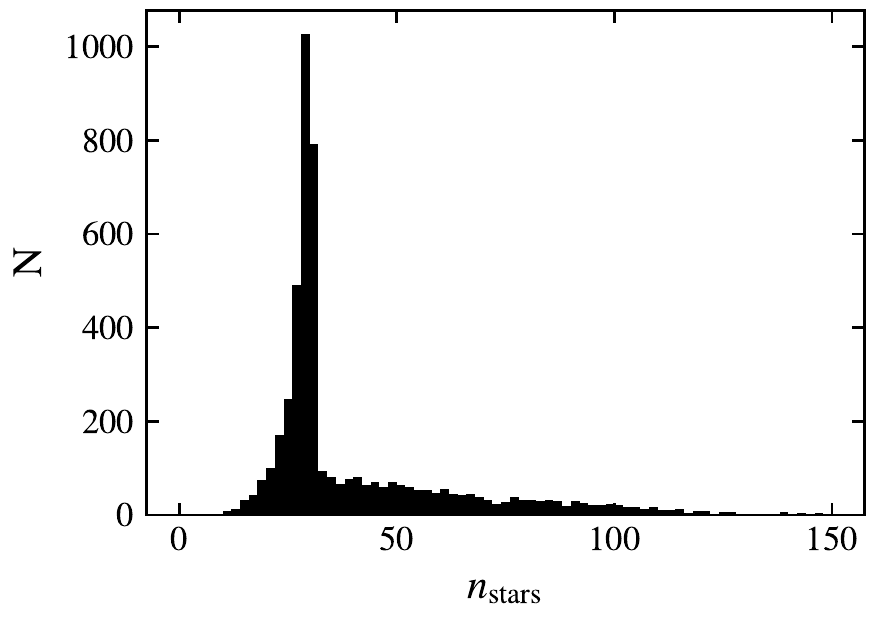}
\caption{Distribution of the number of calibration stars used on each image in
  the survey.  An additional 18 observations have more than 150 calibration stars.}
\label{fig:nstars}
\end{figure}

The pipeline is then run a final time, now applying corrections for both the
residual maps and DCR.  Figure~\ref{fig:rms} shows the distributions of the rms
residuals for the astrometric solutions for all survey images.  The
distributions peak near 4 mas, with 85\% of the solutions having rms values
less than 10 mas.
Figure~\ref{fig:nstars} displays the distribution of the number of calibrating
stars used for each image.  The sharp drop after 30 stars is because the
pipeline only uses stars with better than 0.5\% photometry, but will use less
well-exposed stars if necessary to get at least 30 stars.  Not all fields have
30 stars available, well-exposed or not.

The astrometric solutions were then used to derive ICRS astrometric-place
coordinates for the target star on each image, at the epoch of the observation.
DCR corrections to the target stars were applied using synthesized Pan-STARRS $r-i$ and $i-z$ colors from
the ATLAS All-Sky Stellar Reference Catalog \citep{2018ApJ...867..105T}.
Note that the median parallax (as measured by Hipparcos-2) for stars in the
survey is 10 mas, with nearly a third of the stars having parallaxes greater
than 20 mas, easily detectable at the level of astrometry achieved by the
survey.

\subsection{Comparison with Gaia for Single Observations}

\begin{figure}
\plotone{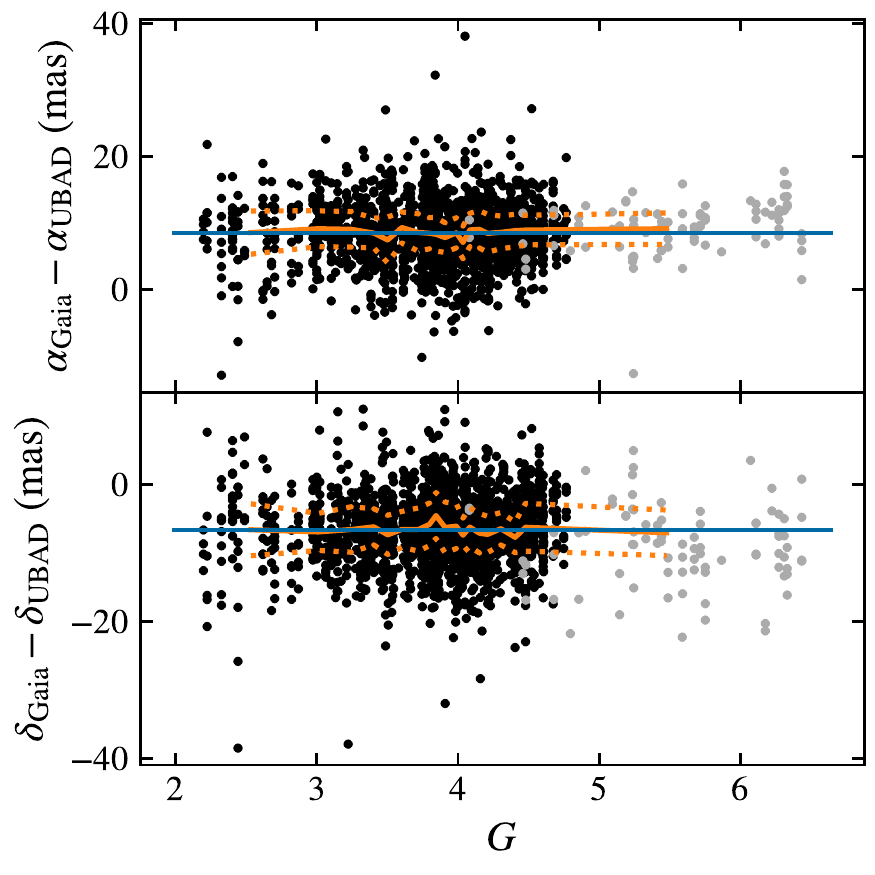}
\caption{Differences between Gaia EDR3 and UBAD individual-frame positions
  versus Gaia $G$ magnitude for a clean set of $i$ observations (right
  ascension in the top panel, declination in the bottom panel).  The black points are observations for the main survey while the gray points are for the faint survey extension, used only in this figure.  The blue line is the median difference for stars in the main survey.  The solid and
  dotted orange lines indicate the median and interquartile range, respectively, of the differences in bins of 100
  stars in $G$.}
\label{fig:g-diff-I}
\end{figure}

\begin{figure}
\plotone{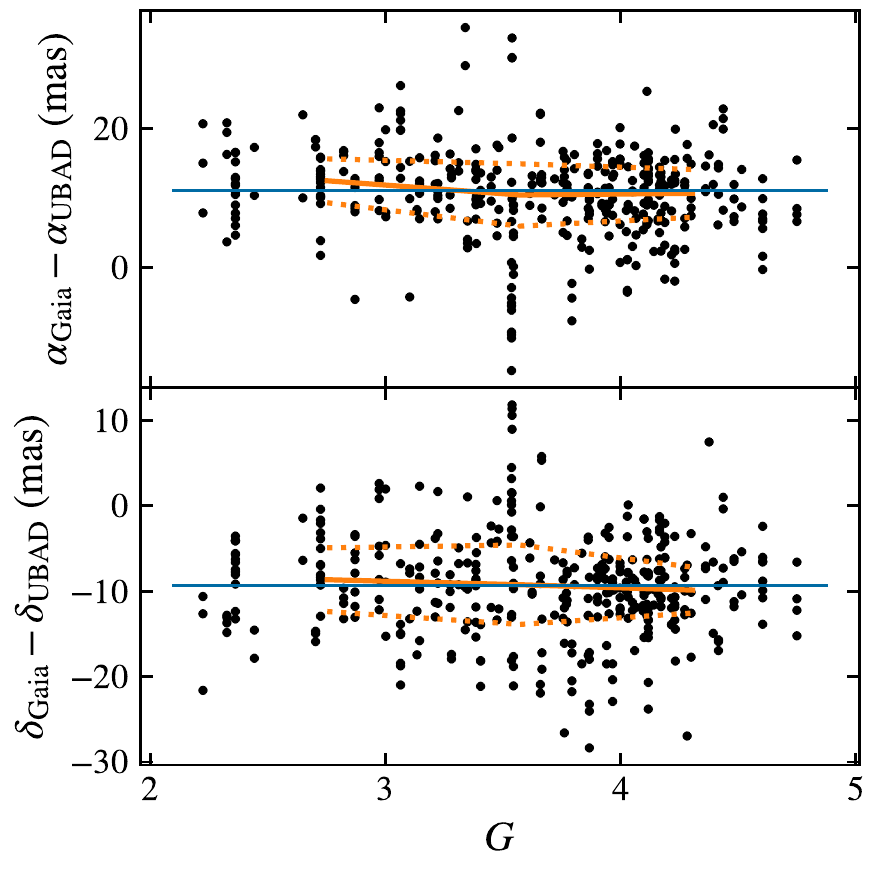}
\caption{Same as Figure~\ref{fig:g-diff-I}, but for $z$ observations.}
\label{fig:g-diff-Z}
\end{figure}

Of the 4958 observations which comprise the UBAD survey, 4607 target a star
which is in Gaia EDR3.  Figures~\ref{fig:g-diff-I} and
\ref{fig:g-diff-Z} plot the differences between the target star ICRS
coordinates measured on each UBAD image and the matching Gaia EDR3 catalog positions,
propagated to astrometric-place coordinates at the epoch of the UBAD observations
using Gaia's proper motions and parallaxes, for the 1998 $i$
and 371 $z$ observations which meet the following criteria:
\begin{enumerate}
\item seeing is less than 2 arcsec;
\item the astrometric calibration uses at least 15 reference Gaia stars (relaxed for a few sparse fields) and
  yields rms residuals of less than 20 mas in both right ascension and
  declination;
\item visual inspection of the target star on the UBAD image indicates no issues (blends, close
  neighbors, etc);
\item Hipparcos-2 flags the star as a single star with a satisfactory
  5-parameter astrometric fit;
\item the matching Gaia star has a 5- or 6-parameter solution and
  $\textbf{ruwe} < 3$.
\end{enumerate}
Black and gray points are for stars in the main survey and the faint extension, respectively
(no $z$ observations were taken for the faint extension). The blue lines are the median differences for stars in the main survey.
The solid and dotted orange lines indicate the medians and interquartile
ranges, respectively, of the differences in bins of 100 stars in $G$.  The
comparison is excellent, with no magnitude-dependent systematic trends
in either the fully-expected median differences (discussed below) or the dispersions over the four magnitudes of overlap in $G$ between UBAD and Gaia.

\begin{figure}
\plotone{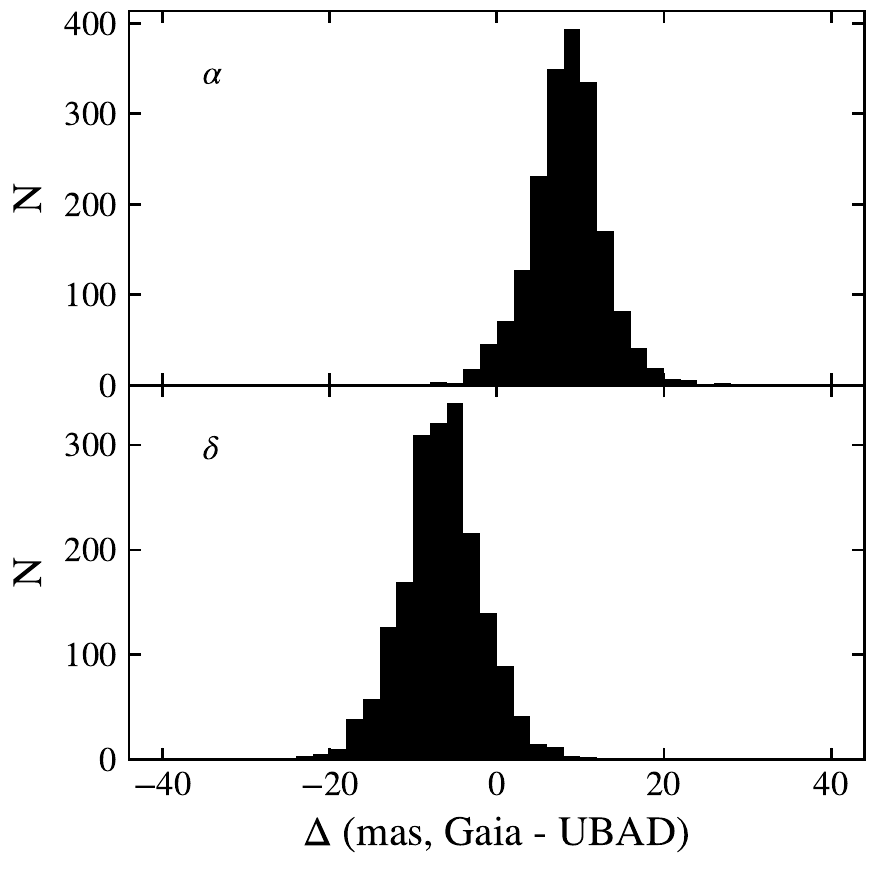}
\caption{Histograms of the differences between Gaia and UBAD positions for a
  clean set of $i$ observations.  The differences in
  right ascension and declination are shown in the top and bottom panels,
  respectively.}
\label{fig:gaia-hist-I}
\end{figure}

\begin{figure}
\plotone{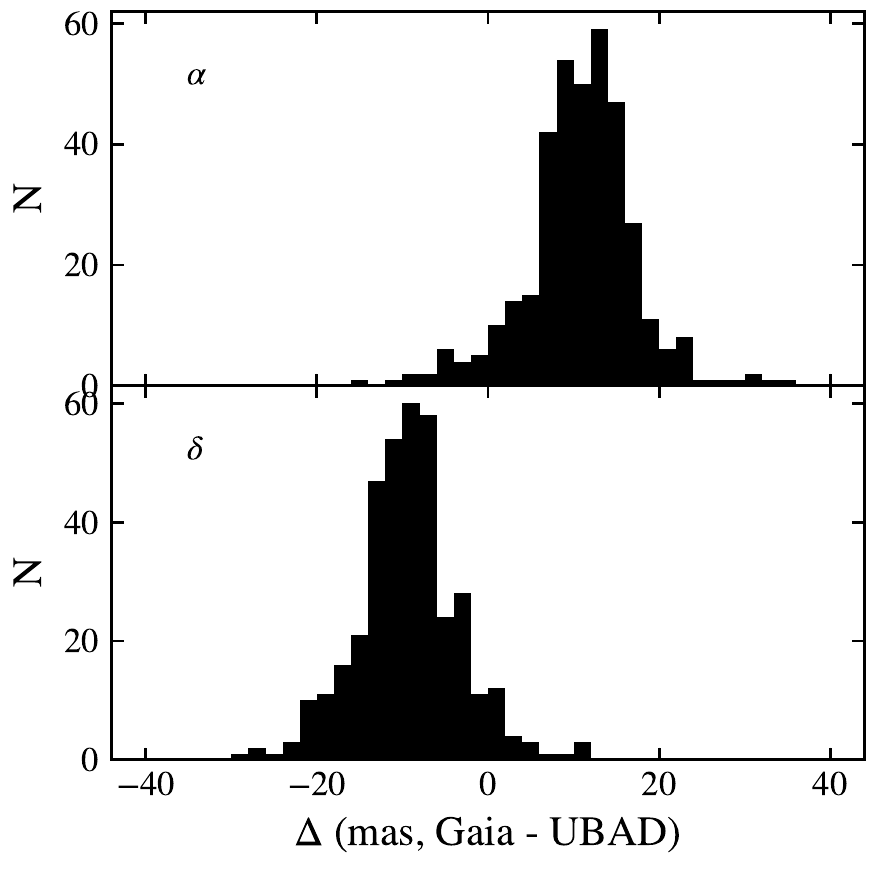}
\caption{Same as Figure~\ref{fig:gaia-hist-I}, but for $z$ observations.}
\label{fig:gaia-hist-Z}
\end{figure}

Figures~\ref{fig:gaia-hist-I} and \ref{fig:gaia-hist-Z} display histograms of
the differences between the UBAD and Gaia EDR3 positions in $i$ and $z$
shown in Figures~\ref{fig:g-diff-I} and \ref{fig:g-diff-Z}, limited to stars in the main survey.  The median
differences (Gaia $-$ UBAD) in $i$ are 8.5 and -6.6 mas in right
ascension and declination, respectively.  In $z$ the median
differences are 11.1 and -9.3 mas in right ascension and declination,
respectively.  Systematic offsets of such size are expected as the
residual map was not generated under the spot, and so these offsets can just be
thought of as part of the residual map.  These median offsets are added to all UBAD
individual observation positions and are applied for all subsequent
analyses.  Note that since we use the offsets to calibrate UBAD, we
cannot validate that Gaia EDR3 itself does not have a magnitude-independent
systematic bias over the magnitude range covered by UBAD, however, given that
there is no magnitude-dependent systematic offset over that magnitude range, Gaia is
likely free of any such overall systematic error.  The rms
differences in $i$ are 4.5 and 5.0 mas in right ascension and
declination, respectively.  In $z$ the rms differences are 6.5 and 6.1
mas in right ascension and declination, respectively, giving a good indication
of the single-frame astrometric accuracy of the UBAD observations (typical errors in the Gaia positions at the UBAD observation epochs are less than a mas).  Different
quality cuts to determine which stars are included in the analysis change the
offsets by of order 0.5 mas in $i$ and 1 -- 2 mas in $z$, which is one
indication of the size of systematic errors in the UBAD astrometry.

\begin{figure}
\plotone{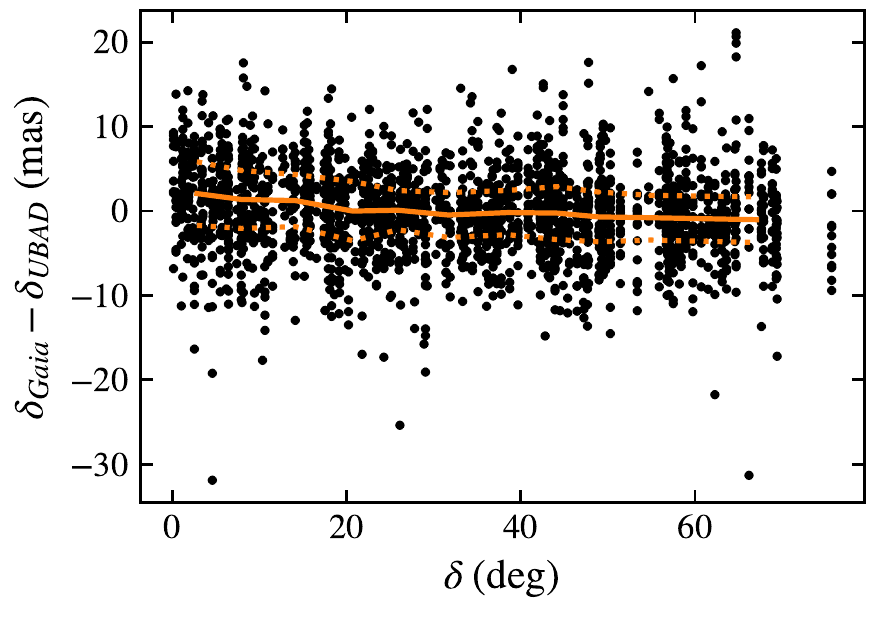}
\caption{Offsets in declination between Gaia and UBAD positions as a function
  of declination, after correction for the median offsets.  The orange solid and
  dashed lines are the median and interquartile range of the offsets,
  respectively, in bins of 200 stars.}
\label{fig:dec-systematics}
\end{figure}

After correcting for the median offsets, there remain systematic differences
between Gaia EDR3 and UBAD declinations as a function of declination.
Figure~\ref{fig:dec-systematics} plots these differences.
The orange solid and dashed lines indicate the median and
interquartile range offsets, respectively, in bins of 200 stars.
No similar systematic differences are seen in right ascension.  We don't
understand the cause of the systematic offsets.  One possible source is the
ring of reflected light around the target star seen in Figure~\ref{fig:spot}.
The star is offset along the x-axis with respect to that ring; the x-axis is
parallel to the declination axis.  It's possible that as the telescope flexes at different
declinations (it is on an equatorial mount), the offset between the star and
the center of the ring changes, affecting the measured center of the star.
The brightness of the ring is only a few percent that of the peak of the star,
however, the systematic offset is small, varying roughly linearly by about 3 mas
over the entire declination range.  Another possible source is DCR perhaps not being
fully corrected by the adopted correction model; most observations were taken within
an hour of the meridian, so refraction is predominantly along declination.
It is far less likely that the systematics are in the Gaia EDR3 positions; Gaia's
scanning law is roughly symmetric in ecliptic coordinates, and without concerns
regarding atmospheric refraction or gravitational flexure, there is unlikely to be any
systematic with declination.  As the systematic is poorly understood and given
that the comparison is with heavily-saturated Gaia stars, we choose not to
correct it but note it as another potential source of systematic error.

\subsection{Estimating Astrometric Errors for Single Observations}

Characterizing the single-frame astrometric errors is complicated as the
target star is observed in an environment (through 12.5 magnitudes of
attenuation and on a background containing reflected and scattered light from
the bright target star) very different from the remaining stars on the image.
The errors are characterized by comparison with Gaia as the Gaia errors
(less than a mas) are much smaller than the UBAD single-frame errors.  Based on
that comparison, it is clear that simply combining the center errors measured
by SExtractor in quadrature with the calibration errors underestimates the true
errors.  A floor of 3.5 mas is imposed
on the center error of the target star on individual frames, which is then
added to the center error measured by SExtractor (typically about one mas).
This is then added in quadrature with the rms of the residuals of the
astrometric calibration divided by the square root of the number of
calibrating stars.  This yields typical errors of around five mas separately in
right ascension and declination, consistent with the histograms of differences
between Gaia and UBAD positions seen in Figures~\ref{fig:gaia-hist-I} and
\ref{fig:gaia-hist-Z}.

\begin{figure}
\plotone{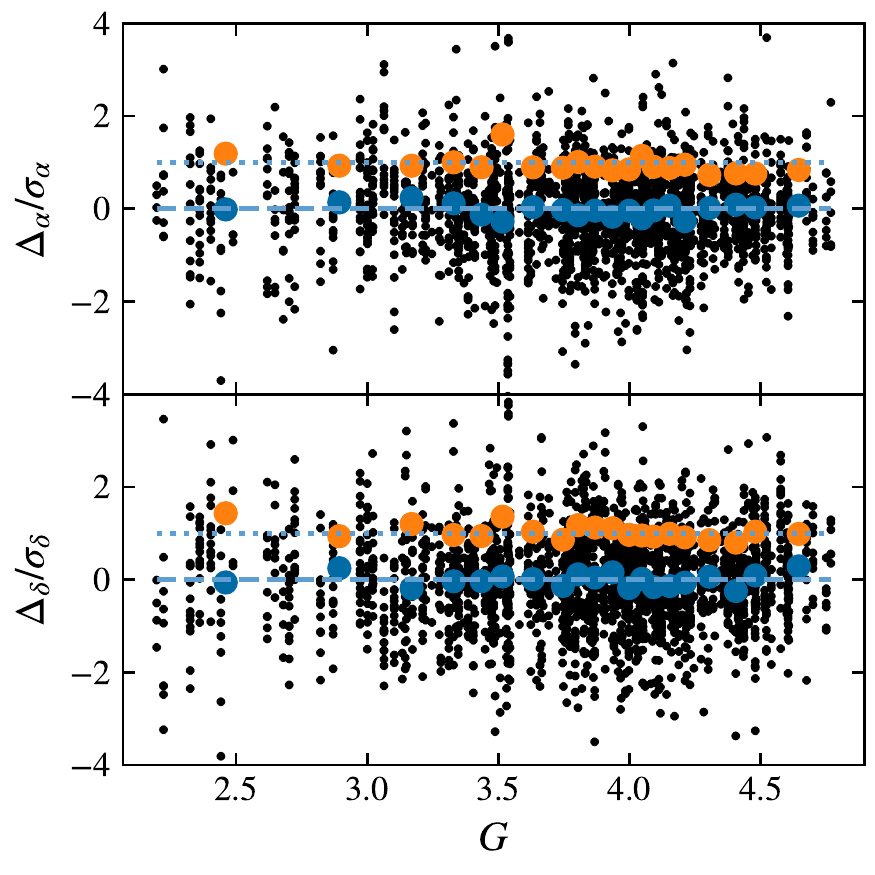}
\caption{Differences between Gaia and UBAD positions (Gaia -- UBAD, right
  ascension in the top panel, declination in the bottom panel), normalized by
  the expected error in the differences.  The black points are the individual
  stars.  The blue and orange points are the mean
  and rms differences, respectively, in equal-sized bins in $G$.  The rms
  values should be near one (indicated by the dotted cyan line).  The dashed cyan line is a reference line at zero normalized error.}
\label{fig:errors}
\end{figure}

Figure~\ref{fig:errors} plots the differences between UBAD and Gaia EDR3 positions,
normalized by the expected errors in the differences based on the modified UBAD
error estimates.  The mean and rms differences are indicated in $G$ magnitude
bins along with the individual star differences.  These normalized differences
should follow a Gaussian distribution with an rms of 1 if the UBAD error
estimates are valid (since the UBAD errors are significantly larger than the
Gaia errors, the UBAD errors dominate this analysis);  that is the case
at all magnitudes.

\section{Catalog Astrometry}
\label{sec:catalog-astrometry}

\subsection{Positiona and Proper Motion Fits}
\label{sec:catalog-astrometry:fits}

\begin{figure*}
\plotone{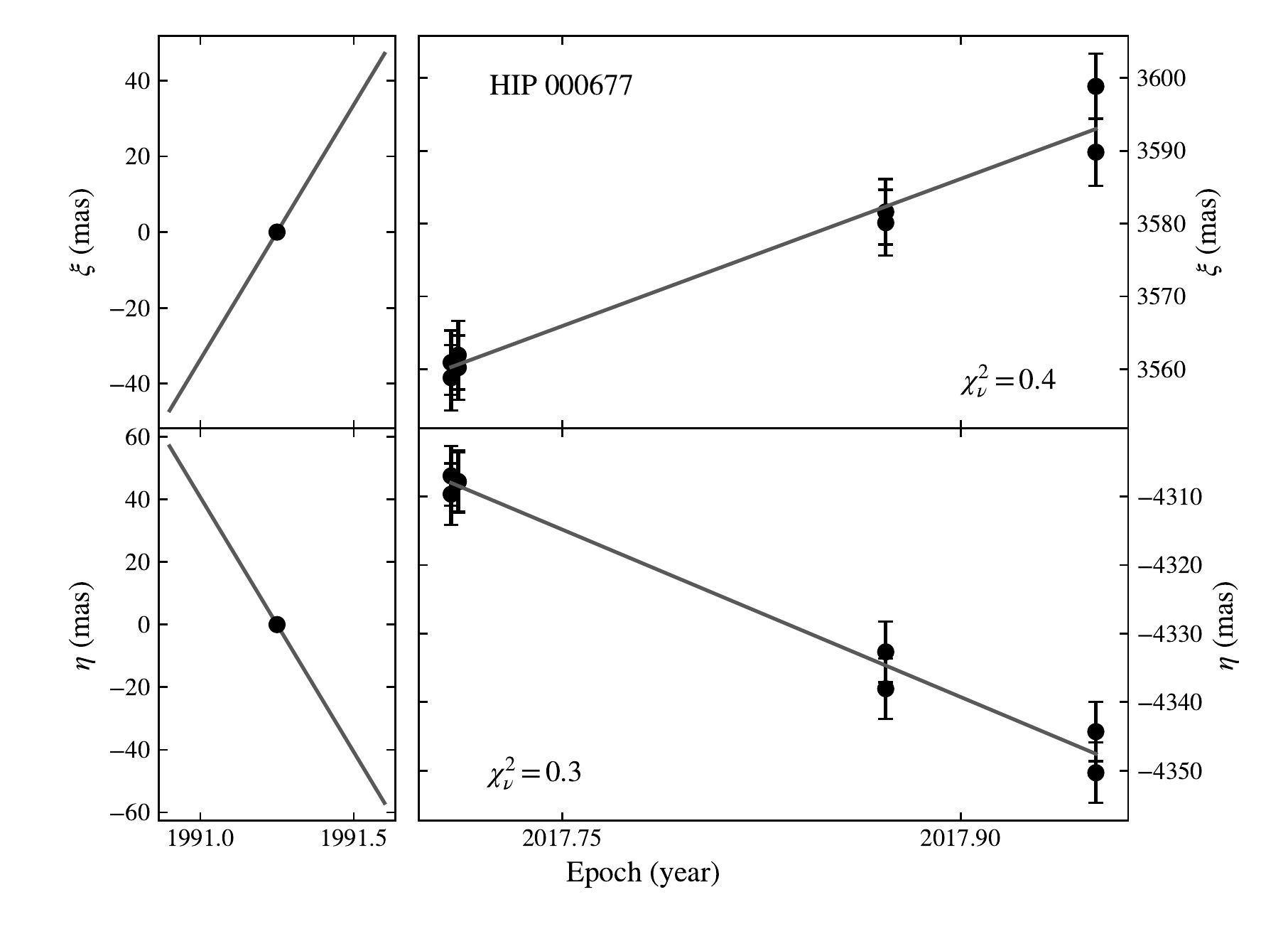}
\caption{Fits to position and proper motion in the tangent plane, $\xi$\ in the top panel, $\eta$\ in the bottom panel, for HIP 000677.  The circles with error bars are the individual observations and the lines are the fits.  The Hipparcos-2 position (left panels) is taken as the tangent point and is plotted separate from the UBAD observations (right panels) due to the large epoch difference.  The complete figure set (364 images) is available in the online journal.}
\label{fig:fits}
\end{figure*}

For each survey star, position and proper motion are fit separately in
right ascension and declination using the individual UBAD observations
combined with the Hipparcos-2 catalog position (with some exceptions, described below).
As we lack enough UBAD observations to fit a parallax, UBAD positions are
corrected for parallax using the Hipparcos-2 parallax.  Only
UBAD observations taken in less than 2 arcsec seeing and with rms residuals in
both right ascension and declination of less than 20 mas are used.  All stars
have at least four UBAD observations.
Individual-frame positions are weighted by their inverse variance, and a linear fit is performed of position versus epoch.
Since the Hipparcos-2 position errors are less than a mas while the UBAD errors
for the restricted set of observations are all around 5 mas, this is
essentially an unweighted fit to the UBAD positions with the Hipparcos-2
position as a fixed point in the fit.  The fits are displayed in
Figure~\ref{fig:fits}.

\figsetstart
\figsetnum{15}
\figsettitle{Position and Proper Motion Fits}
\include{fits.tex}
\figsetend

Sixty two of the stars have fits in right ascension and/or declination with
$\chi^2_\nu$ values greater than 2.  Thirty three of these stars were refit without
the Hipparcos-2 position, leading to much better fits; most are known
or suspected multiple systems.  These include the following stars:

\begin{enumerate}
\item \textbf{HIP 028734, HIP 043109, HIP 054061, HIP 083895, HIP 088601, HIP 101769, HIP 104887}:
Orbital binaries, fit with 7 parameters in Hipparcos-2 after taking into account the orbital parameters.
\item \textbf{HIP 014328, HIP 037279, HIP 067927, HIP 081693, HIP 095501, HIP 112158}:
Orbital binaries, fit with 5 parameters in Hipparcos-2 after taking into account the orbital parameters.
\item \textbf{HIP 003092, HIP 029655,  HIP 031681, HIP 086974, HIP 099874, HIP 116727}:
Orbital binaries in the Sixth Catalog of Orbits of Visual Binary Stars \citep[ORB6,][]{2001AJ....122.3472H}, though orbital parameters were not taken into account in the Hipparcos-2 solutions.
\item \textbf{HIP 015900, HIP 028358, HIP 100587, HP 111022}:
Spectroscopic binaries in the Ninth Catalog of Spectroscopic Binary Orbits \citep[SB9,][]{2004A&A...424..727P}.
\item \textbf{HIP 017358, HIP 072105:}
  Unresolved doubles in Hipparcos-2, fit using standard 5-parameter solutions.
  HIP 072105 is clearly an unresolved double in the UBAD reductions.
\item \textbf{HIP 109427:}
  The Hipparcos-1 fit indicates signs of accelerated motion, while the Hipparcos-2 fit is marked as stochastic.  A low-mass companion was recently detected \citep{2021AJ....162...44S}.
\item \textbf{HIP 014668, HIP 058948:}
  Astrometric binaries in the $\Delta \mu$ catalog of \citet{2005AJ....129.2420M}.
\item \textbf{HIP 067627, HIP 079804}
  Unresolved photometric variable stars in Hipparcos-2.
\item \textbf{HIP 046853, HIP 102488:}
  These are in the Washington Double Star Catalog \citep[WDS,][]{2001AJ....122.3466M} but not marked as suspect in Hipparcos-2 or listed in the $\Delta \mu$ catalog.
\item \textbf{HIP 101810:}
   No indication of possible binarity.
\end{enumerate}

The remaining 29 stars with $\chi^2_\nu$ values greater than 2 were processed
including the Hipparcos-2 position, as excluding the Hipparcos-2 position did
not significantly decrease the fit $\chi^2_\nu$ values.  These include:
\begin{enumerate}
\item \textbf{HIP 013531, HIP 023453, HIP 080816}:
Orbital binaries, fit with 5 parameters in Hipparcos-2 after taking into account the orbital parameters.
\item \textbf{HIP 004427, HIP 084345, HIP 095947, HIP 097365}:
Orbital binaries in ORB6, though orbital parameters were not taken into account in the Hipparcos-2 solutions.
\item \textbf{HIP 036284, HIP 104732}:
Spectroscopic binaries in DB9.
\item \textbf{HIP 012706:}
  Unresolved double in Hipparcos-2, fit using a standard 5-parameter solution.
\item \textbf{HIP 094648:}
  Astrometric binary in the $\Delta \mu$ catalog.
\item \textbf{HIP 022730, HIP 027989, HIP 091262, HIP 107259}
  Unresolved photometric variable stars in Hipparcos-2.
\item \textbf{HIP 050372:}
  Stochastic fit in Hipparcos-2.
\item \textbf{HIP 003179, HIP 049637, HIP 077070, HIP 084379, HIP 087833, HIP 097278, HIP 112724:}
  These are in the WDS but not marked as
  suspect in Hipparcos-2 or listed in the $\Delta \mu$ catalog.
\item \textbf{HIP 054539, HIP 057399, HIP 083000, HIP 085693, HIP 086742, HIP 117863:}
  These six stars are not known or suspected doubles.  Their fits have
  $\chi^2_\nu$ values in right ascension/declination of
  2.6/1.0, 1.0/2.5, 3.6/1.7, 0.5/3.4, 2.2/5.9, and 2.1/0.9,
  respectively.  The fits all look reasonable, and 
  may just represent the tail of the $\chi^2_\nu$ distributions.
\end{enumerate}

\begin{figure}
\plotone{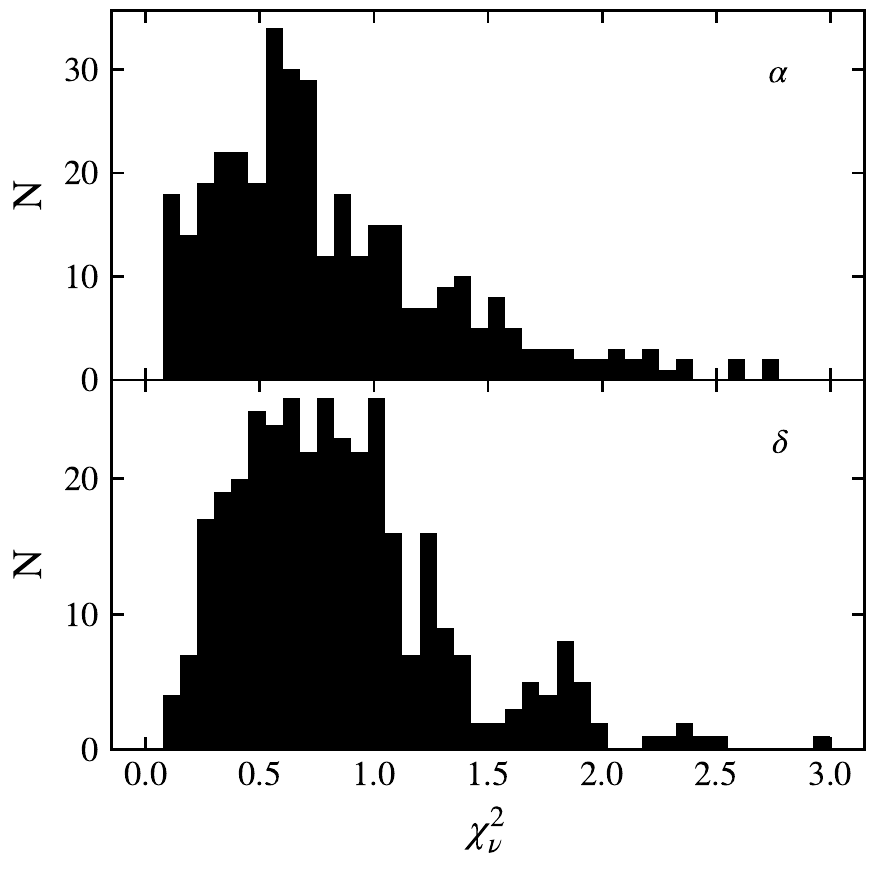}
\caption{$\chi^2_\nu$ distributions for the position and proper motion fits
  in right ascension (top panel) and declination (bottom panel).  There are an
  additional 6 and 10 unplotted stars with $\chi^2_\nu > 3$ in right ascension
  and declination, respectively.}
\label{fig:chi2}
\end{figure}

 Figure~\ref{fig:chi2} displays the distribution of
$\chi^2_\nu$ values from the fits.  These fits yield the astrometric parameters
given in the UBAD catalog.  All UBAD catalog positions are for the epoch
2017.0, roughly the median epoch of the UBAD observations.

\subsection{Comparison of Catalog Astrometry with Gaia and Hipparcos-2}

\begin{figure}
\plotone{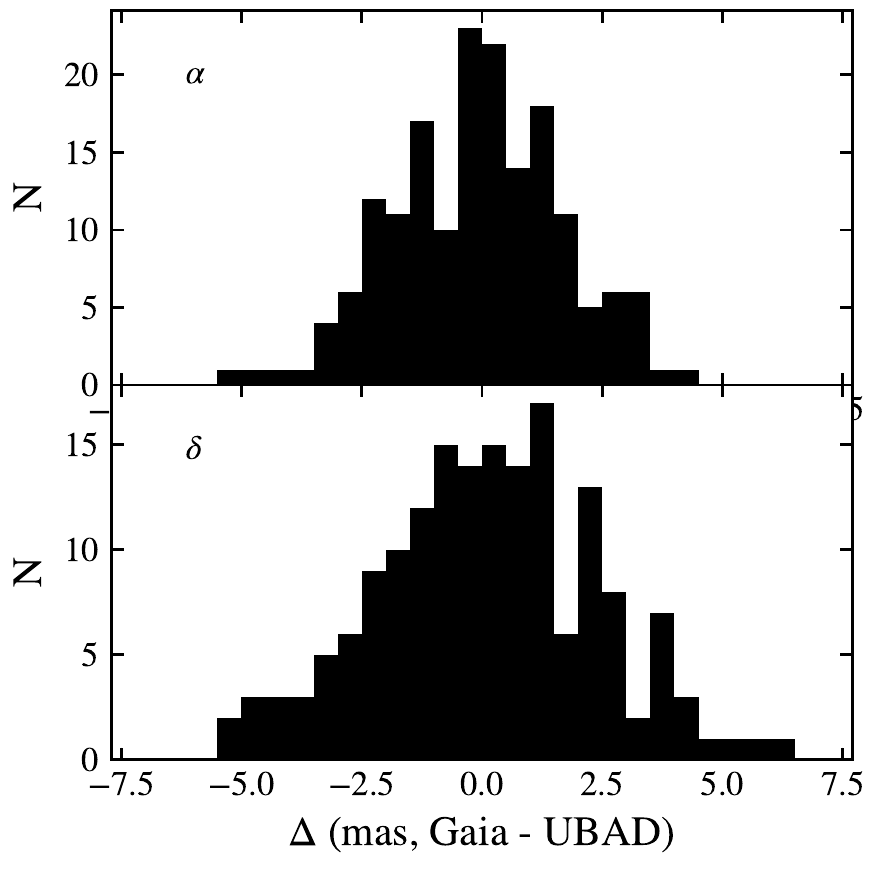}
\caption{Distributions of differences in the catalog positions between Gaia and
  UBAD in right ascension (top panel) and declination (bottom panel).}
\label{fig:stars-diff-hist}
\end{figure}

Figure~\ref{fig:stars-diff-hist} plots the
distributions of the differences between Gaia EDR3 and UBAD catalog positions,
after propagating the Gaia positions to the UBAD catalog epoch.
The sample is limited to stars with good fits in both UBAD
($\chi_\nu^2 < 2$ in both right ascension and declination) and
Gaia (5- or 6-parameter fits and $\textbf{ruwe} < 3$), and that were fit
as single stars in Hipparcos-2.
The median offsets are -0.05 and 0.17 mas in right ascension and declination,
respectively, with rms scatters of 1.8 and 2.3 mas.

The median error for the Gaia positions for the stars in Figure~\ref{fig:stars-diff-hist}, propagated to the UBAD catalog epoch, is around 0.2~mas, with 90\% of the stars having errors less than around 0.35~mas.  The typical errors for the UBAD stars are around 2~mas, consistent with the scatter in Figure~\ref{fig:stars-diff-hist}.  Thus, with position errors approximately 10 times as large as those of Gaia, UBAD can't validate Gaia bright-star astrometry to the stated accuracy in the Gaia catalog.  However, UBAD does confirm Gaia's
bright-star astrometry at approximately the 2~mas level, providing external validation of Gaia's astrometry for these heavily-saturated stars in their survey.

\begin{figure}
\plotone{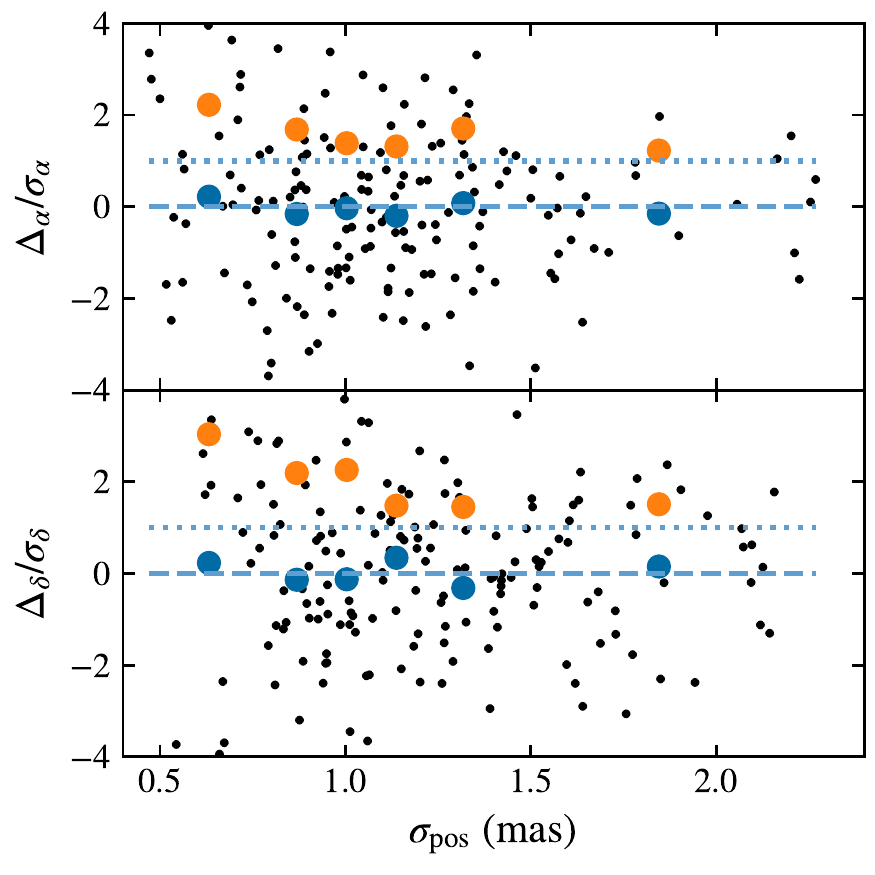}
\caption{Differences between Gaia and UBAD catalog positions (Gaia -- UBAD,
  right ascension in the top panel, declination in the bottom panel),
  normalized by the expected error in the differences, plotted against the
  formal UBAD position error.  The black points are the differences for individual stars.  The blue and orange points are
  the mean and rms differences, respectively, in equal-sized bins.  If the
  error estimates are correct, the rms values should be near one (indicated by
  the dotted cyan line).  The dashed cyan line is a reference line at zero normalized error.}
\label{fig:stars-errors-before}
\end{figure}

\begin{figure}
\plotone{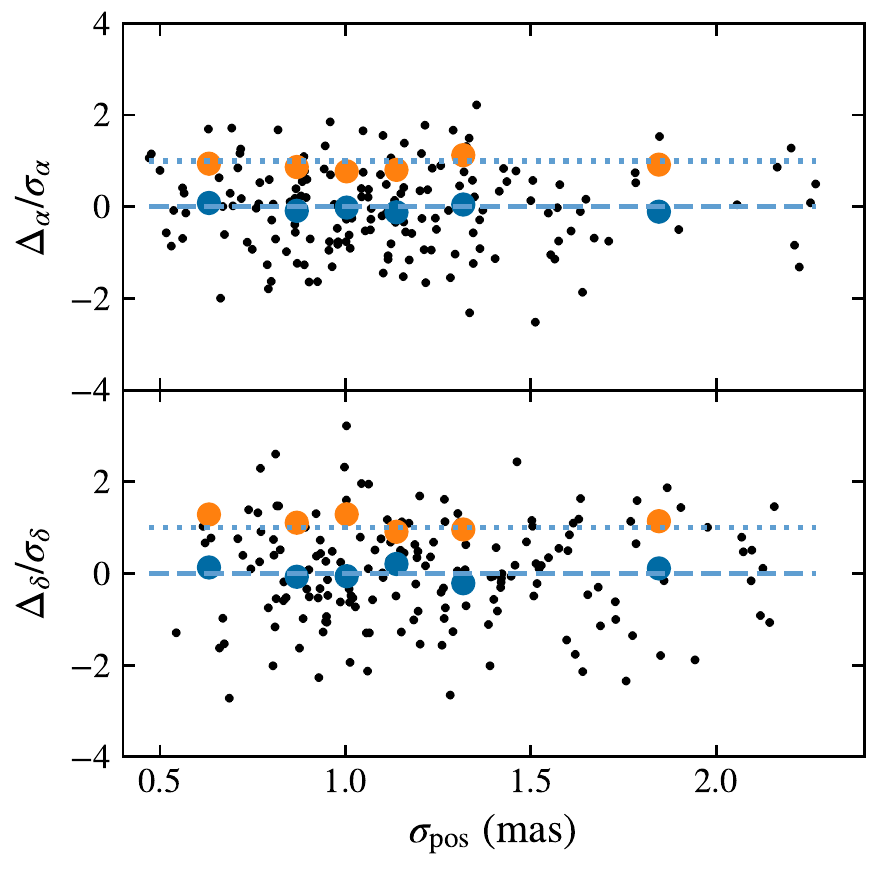}
\caption{Same as Figure~\ref{fig:stars-errors-before}, except that the UBAD
  position errors used to normalize the differences now include a floor error
  of 1.5~mas added in quadrature with the formal error.  The x-axis remains the
  formal error, without the addition of the floor error.}
\label{fig:stars-errors-after}
\end{figure}

The formal errors in the UBAD positions appear to underestimate the true
errors.  Figure~\ref{fig:stars-errors-before} plots the differences between the
Gaia and UBAD catalog positions, normalized by the expected errors in the
differences.  The rms differences in equal-sized bins of the formal UBAD position error are indicated by the orange points.
If the error estimates are correct, the rms values should be near one.  Clearly,
either the UBAD or Gaia position errors are underestimated.  The Gaia errors would have to be underestimated by about a factor of eight to explain the discrepancy, which is unlikely.
Thus, the final UBAD  catalog position errors are corrected by
adding a floor error of 1.5~mas in quadrature to the formal errors.
Figure~\ref{fig:stars-errors-after} repeats
Figure~\ref{fig:stars-errors-before}, but now using the inflated error
estimates; the rms of the position differences now match the expected value of
one for all bins in formal error.

\begin{figure}
\plotone{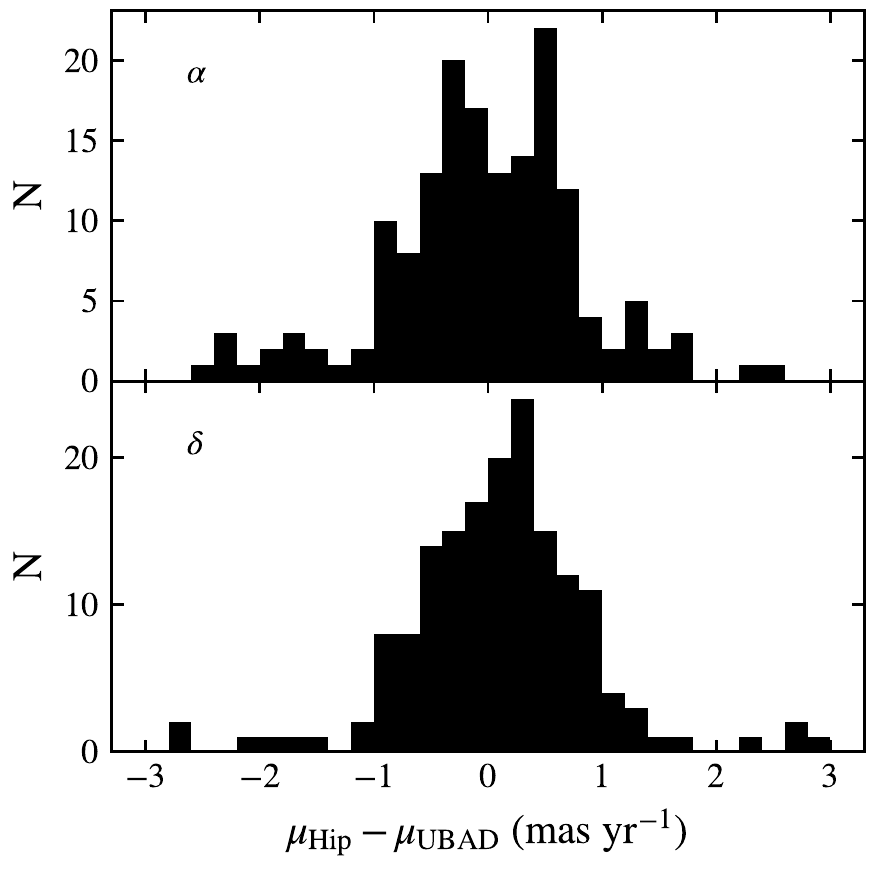}
\caption{Distribution of the differences between Hipparcos-2 and UBAD proper
  motions for a clean sample of stars.  There are an additional nine and six
  stars outside the histogram limits in right ascension and declination,
  respectively.}
\label{fig:pm-hist-hip}
\end{figure}

\begin{figure}
\plotone{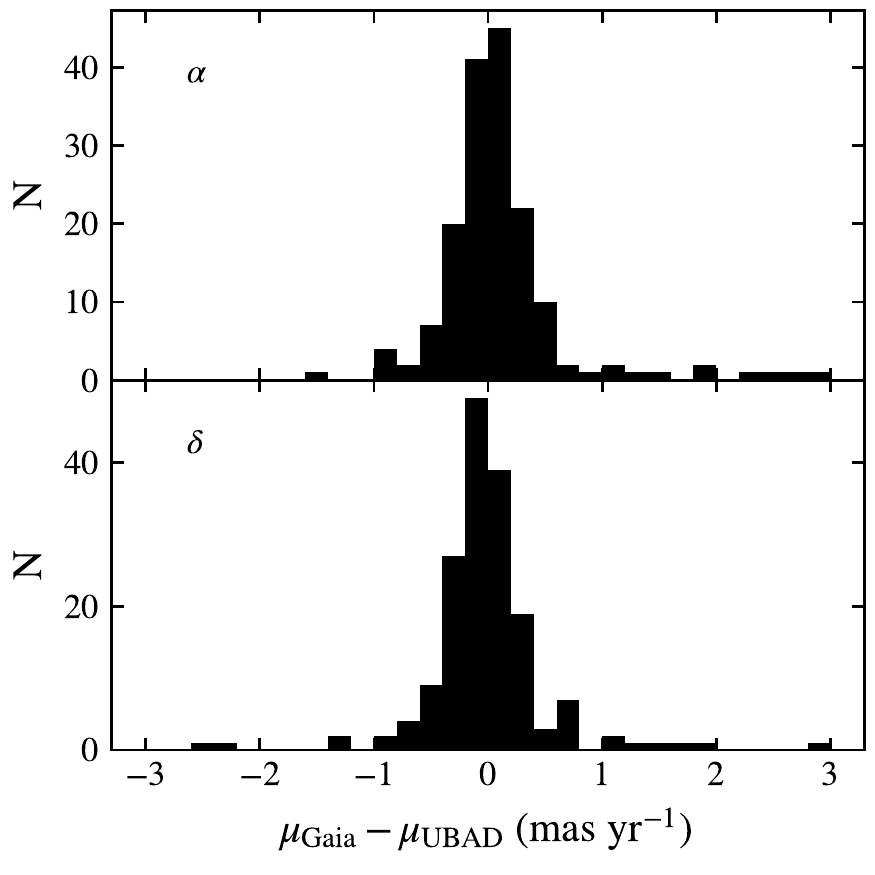}
\caption{Same as Figure~\ref{fig:pm-hist-hip}, but comparing UBAD and Gaia
  proper motions.  There are an additional six and one stars outside the histogram
  limits in right ascension and declination, respectively.}
\label{fig:pm-hist-gaia}
\end{figure}

Figures~\ref{fig:pm-hist-hip} and \ref{fig:pm-hist-gaia} display the
distributions of the differences between the UBAD proper motions and those of
Hipparcos-2 and Gaia EDR3, respectively, limited to the same clean sample used in
Figure~\ref{fig:stars-diff-hist}.  There are no significant systematic offsets
between the catalogs.  The formal proper motion errors for UBAD are smaller
than those for Gaia by typical factors of two to six and considerably smaller
than those for Hipparcos-2, benefiting from the roughly 26-year epoch difference
between Hipparcos and UBAD (clearly one could combine Gaia EDR3 and Hipparcos-2 positions to derive proper motions more accurate than UBAD's motions). Given the underestimated UBAD position errors, it's
likely that the UBAD proper motions errors are also underestimated.

\section{The Catalog}
\label{sec:catalog}

The UBAD catalog is presented in Table~\ref{tab:catalog}.  The catalog contains 364 stars, 36 of which are not included in Gaia EDR3.  All catalog positions are given for the epoch 2017.0, roughly the median epoch of the observations.

\begin{splitdeluxetable*}{rrrrrrrBrrrrrrrrrr}
\tablecaption{UBAD Catalog\label{tab:catalog}}
\tablehead{
   \colhead{Hip}
 & \dcolhead{\alpha}
 & \dcolhead{\sigma_\alpha\ \textrm{cos}\ \delta}
 & \dcolhead{\delta}
 & \dcolhead{\sigma_\delta}
 & \dcolhead{\mu_\alpha\ \textrm{cos}\ \delta}
 & \dcolhead{\sigma_{\mu_\alpha\ \textrm{cos}\ \delta}}
 & \dcolhead{\mu_\delta}
 & \dcolhead{\sigma_{\mu_\delta}}
 & \dcolhead{N_i}
 & \dcolhead{N_z}
 & \colhead{Hip\_used}
 & \colhead{RMS$_\alpha$}
 & \dcolhead{\chi^2_{\nu,\alpha}}
 & \colhead{RMS$_\delta$}
 & \dcolhead{\chi^2_{\nu,\delta}}
 & \colhead{Class}
 \\
   \colhead{}
 & \colhead{(deg)}
 & \colhead{(mas)}
 & \colhead{(deg)}
 & \colhead{(mas)}
 & \colhead{(mas year$^{-1}$)}
 & \colhead{(mas year$^{-1}$)}
 & \colhead{(mas year$^{-1}$)}
 & \colhead{(mas year$^{-1}$)}
 & \colhead{}
 & \colhead{}
 & \colhead{}
 & \colhead{(mas)}
 & \colhead{}
 & \colhead{(mas)}
 & \colhead{}
 & \colhead{}
}
\colnumbers
\startdata
   677 &   2.0976353 &  1.78 &  29.0896637 &  1.70 &   134.564 &  0.0369 &  -162.824 &  0.0314 &  8 &  0 & 1 &  2.60 & 0.378 &  2.18 & 0.282 & s \\
   746 &   2.2993446 &  1.74 &  59.1489328 &  1.68 &   523.668 &  0.0339 &  -179.656 &  0.0293 & 12 &  0 & 1 &  2.95 & 0.481 &  2.58 & 0.365 & s \\
  1067 &   3.3089709 &  1.98 &  15.1835486 &  1.83 &     1.681 &  0.0499 &    -9.443 &  0.0410 & 11 &  0 & 1 &  4.14 & 0.756 &  3.77 & 0.571 & s \\
  1168 &   3.6511466 &  1.65 &  20.2067096 &  2.45 &    91.341 &  0.0271 &     1.746 &  0.0754 &  9 &  0 & 1 &  1.89 & 0.165 &  5.79 & 1.440 & s \\
  2219 &   7.0127137 &  2.12 &  17.8932221 &  2.42 &   114.618 &  0.0583 &    20.559 &  0.0740 & 12 &  0 & 1 &  5.29 & 1.139 &  6.85 & 1.931 & s \\
\enddata
\tablecomments{Column (1) is the Hipparcos identifier.  Columns (2) -- (5) are the ICRS catalog mean place coordinates and their errors at the catalog epoch of J2017.  Columns (6) -- (9) are the proper motions and their errors.  Columns (10) and (11) are the number of UBAD $i$ and $z$ observations used in the fits, respectively.  Column (12) indicates whether the Hipparcos-2 coordinate was used in the fits (1=yes, 2=no).  Columns (13) -- (16) are the rms of the residuals for the UBAD observations only and the reduced $\chi^2$ of the fits.  Column (17) is the visual classification of the target star on the UBAD images (s=single, n=resolved neighbor, r=resolved blend, u=unresolved blend).  Table~\ref{tab:catalog} is published in its entirety in the machine-readable format.  A portion is shown here for guidance regarding its form and content.}
\end{splitdeluxetable*}

Table~\ref{tab:observations} lists the positions measured on each of the individual images comprising the survey.  Positions are given as ICRS astrometric-place coordinates at the epoch of the observation (uncorrected for parallax and proper motion).  Not all positions in this Table were used in compiling the final catalog (see Section~\ref{sec:catalog-astrometry:fits}).

\begin{splitdeluxetable*}{rrrrrrBrrrrrrr}
\tablecaption{Single-epoch Positions\label{tab:observations}}
\tablehead{
   \colhead{Epoch}
 & \colhead{Hip}
 & \dcolhead{\alpha}
 & \dcolhead{\sigma_\alpha\ \textrm{cos}\ \delta}
 & \dcolhead{\delta}
 & \dcolhead{\sigma_\delta}
 & \colhead{T}
 & \colhead{X}
 & \dcolhead{N_s}
 & \colhead{RMS$_\alpha$}
 & \colhead{RMS$_\delta$}
 & \colhead{Seeing}
 & \colhead{Elong}
 \\
   \colhead{(year)}
 & \colhead{}
 & \colhead{(deg)}
 & \colhead{(mas)}
 & \colhead{(deg)}
 & \colhead{(mas)}
 & \colhead{(s)}
 & \colhead{}
 & \colhead{}
 & \colhead{(mas)}
 & \colhead{(mas)}
 & \colhead{(arcsec)}
 & \colhead{}
}
\colnumbers
\startdata
2016.190580 &  26727 &  85.1897079 &  6.75 & -1.9425936 &  5.35 &  180.0 & 1.279 &  27 &  23.35 &   9.87 & 1.457 & 1.057 \\
2016.190593 &  26727 &  85.1897094 &  5.69 & -1.9425971 &  5.35 &  360.0 & 1.288 &  27 &  16.50 &  13.13 & 1.604 & 1.035 \\
2016.190680 &  34912 & 108.3475446 &  4.62 & 51.4288135 &  4.62 &  750.0 & 1.042 &  35 &   3.37 &   3.31 & 1.259 & 1.015 \\
2016.190704 &  34912 & 108.3475466 &  4.62 & 51.4288122 &  4.59 &  600.0 & 1.043 &  35 &   2.52 &   3.00 & 1.177 & 1.039 \\
2016.190761 &  37946 & 116.6638232 &  4.83 & 37.5174545 &  4.75 &  360.0 & 1.002 &  30 &   3.72 &   3.54 & 1.155 & 1.074 \\
\enddata
\tablecomments{Column (1) is the observation epoch.  Column (2) is the Hipparcos identifier.  Columns (3) -- (6) are the ICRS astrometric place coordinates and their errors at the observation epoch.  Columns (7) -- (8) are the observation exposure time and weighted mean airmass, respectively.  Column (9) is the number of Gaia EDR3 reference stars used in the plate solution.  Columns (10) -- (11) are the rms residuals of the plate solution.  Columns (12) and (13) are the mean FWHM and elongation of stars on the images, respectively.  Table~\ref{tab:observations} is published in its entirety in the machine-readable format.  A portion is shown here for guidance regarding its form and content.}
\end{splitdeluxetable*}

\begin{figure}
\plotone{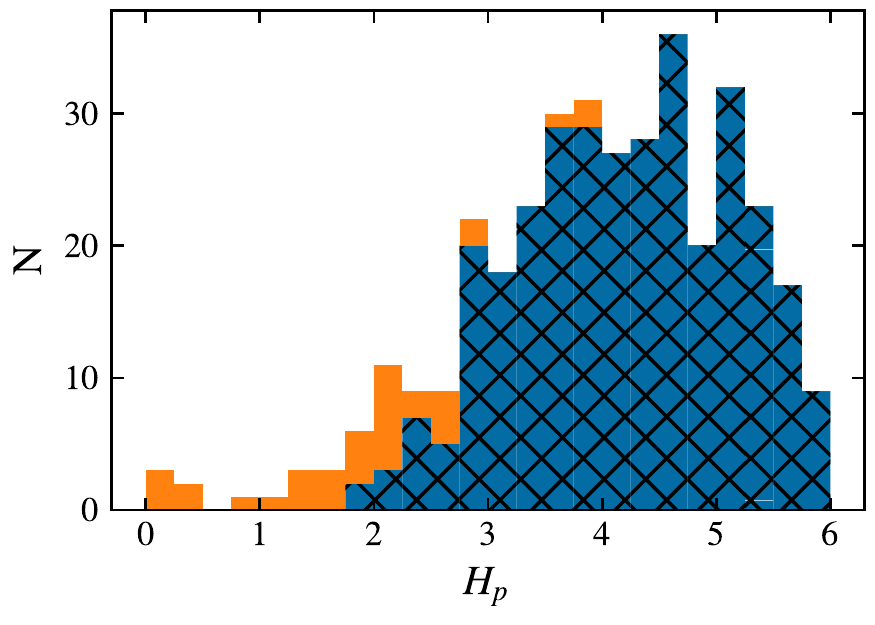}
\caption{Stacked histograms of catalog stars in Hipparcos-2 $H_p$ magnitude.  The
  hatched blue and unhatched orange histograms are for stars with and without an entry in Gaia EDR3, respectively.}
\label{fig:hp-hist}
\end{figure}

\begin{figure}
\plotone{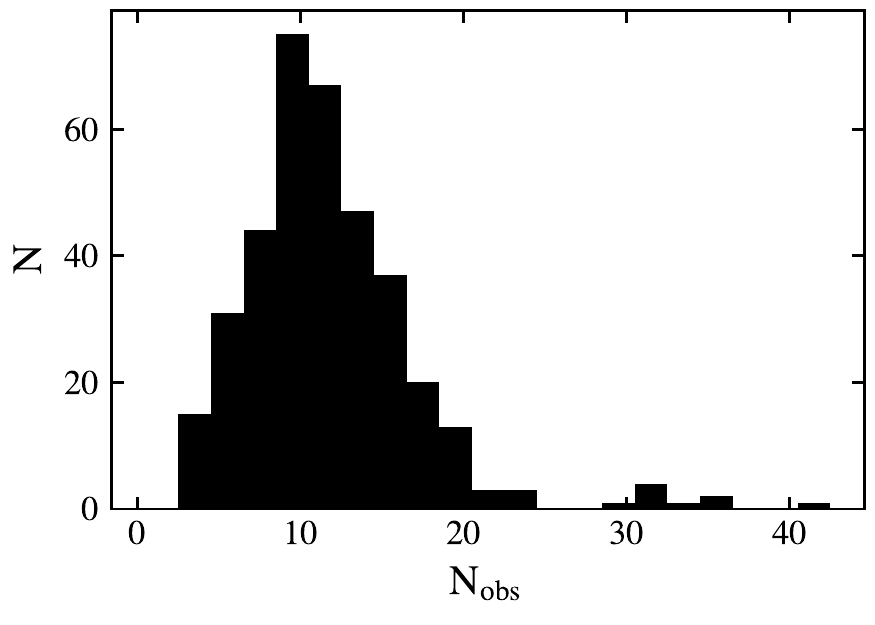}
\caption{Distribution of the number of UBAD observations used in the astrometry
  fits.}
\label{fig:nobs}
\end{figure}

\begin{figure}
\plotone{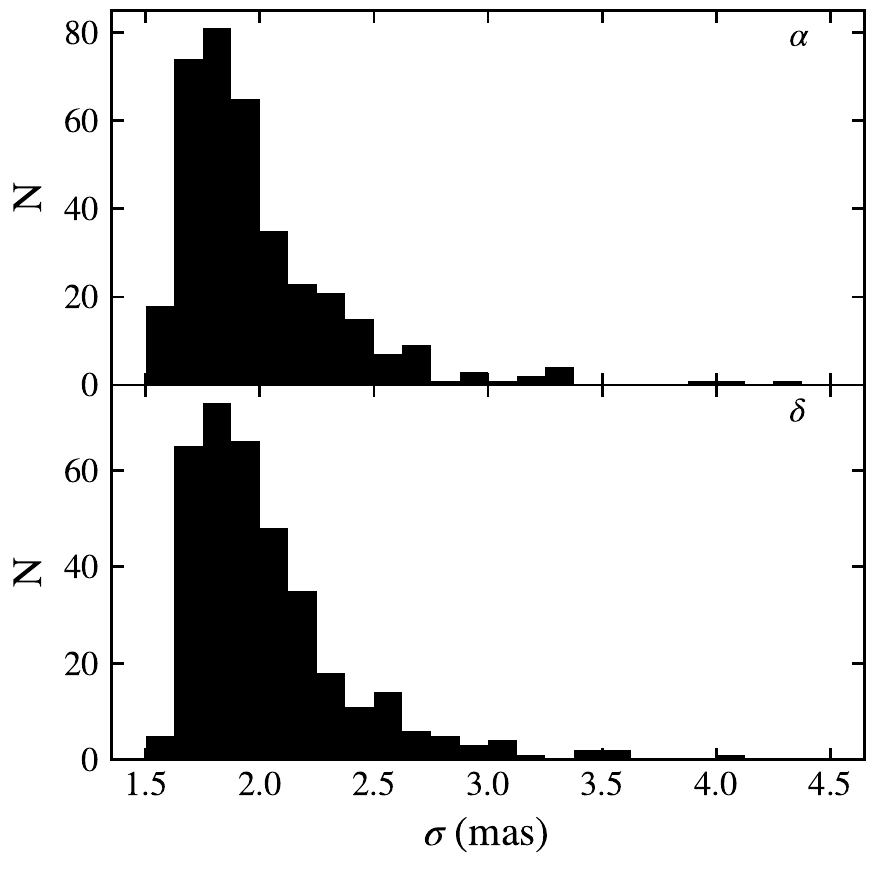}
\caption{Distributions of catalog position errors (right ascension in top
  panel, declination in bottom panel).  There are an additional two and four unplotted stars with errors greater than 4.5 mas in right ascension and declination, respectively.}
\label{fig:pos-errors-hist}
\end{figure}

\begin{figure}
\plotone{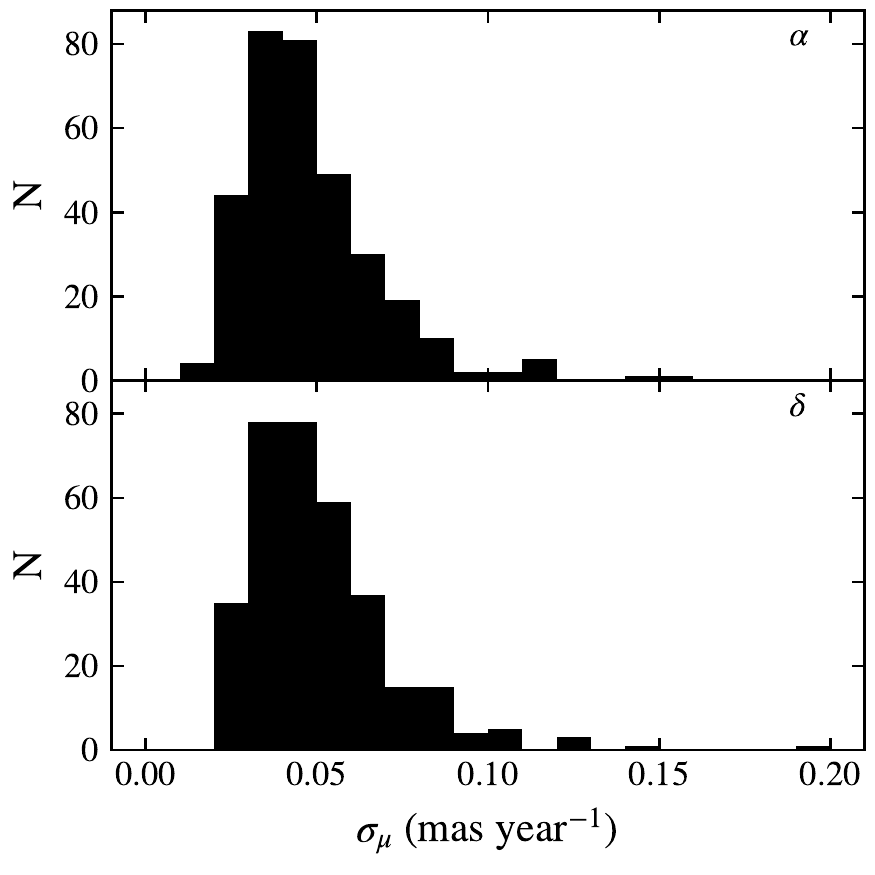}
\caption{Distributions of catalog proper motion errors for the 331 stars whose fits include the Hipparcos-2 position (right ascension in top panel, declination in bottom panel).}
\label{fig:pm-errors-hist-hip}
\end{figure}

\begin{figure}
\plotone{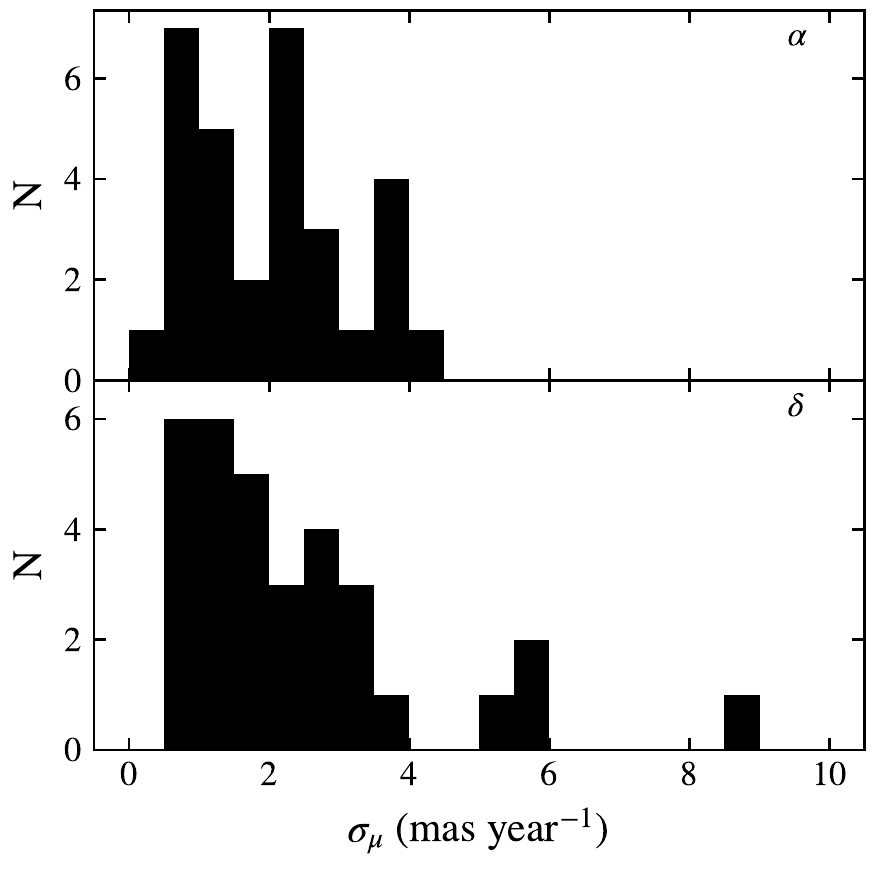}
\caption{Same as Figure~\ref{fig:pm-errors-hist-hip}, but for the 33 stars whose fits exclude the Hipparcos-2 position.  There are an additional two and one unplotted stars with errors greater than 10 mas~year$^{-1}$ in right ascension and declination, respectively.}
\label{fig:pm-errors-hist-nohip}
\end{figure}

Figures~\ref{fig:hp-hist},
\ref{fig:nobs}, \ref{fig:pos-errors-hist}, \ref{fig:pm-errors-hist-hip}, and \ref{fig:pm-errors-hist-nohip} display
the distributions of the catalog stars in Hipparcos-2 $H_p$ magnitude, number of
UBAD observations used in the fits, position errors (at the catalog epoch of
2017.0), and proper motion errors for those stars which included and excluded the Hipparcos-2 position, respectively.  The median position errors
are 1.9 mas in both right ascension and declination; 90\% of catalog stars have
position errors of less than 2.6 mas in either coordinate.  We estimate
systematic errors in the positions of order 1 -- 3 mas.  For the 331 catalog stars whose fits include the Hipparcos-2 position, the median proper motion
errors are 0.044 and 0.047~mas~year$^{-1}$ in right ascension and declination,
respectively, with 90\% of such stars having proper motion errors less than
0.075~mas~year$^{-1}$.  Those stars whose fits excluded the Hipparcos-2 position have proper motion errors of order a few mas~year$^{-1}$, depending on the time span of the UBAD observations for each star.  Systematic errors in the proper motions are harder to
estimate, as no bright-star catalog in this magnitude range has at least
formally better errors in proper motions than UBAD, due to the long time span
by including Hipparcos-2 positions in the proper motion calculations (though
combining Hipparcos-2 with Gaia EDR3 would yield such a catalog).  The
systematic errors in the positions alone would imply systematic errors in
proper motions of order 0.1~mas~year$^{-1}$.

\section{Conclusion}
\label{sec:conclusion}

Gaia has revolutionized astrometry.  While not designed for bright-star astrometry, Gaia EDR3 includes all but a few dozen of the brightest stars.  Stars brighter than $G \sim 5$ are saturated on the Gaia detectors, and the brightest stars are heavily saturated.  Given the importance of such stars to government and commercial interests, it is important to provide some external validation for these stars.  UBAD provides such validation at the few milliarcsecond level, as well as the most accurate current-epoch astrometry for those bright northern hemisphere stars excluded from the current Gaia catalog.

UBAD gives precise positions and proper motions for 364 bright northern-hemisphere stars, including all but five such stars with either $V < 3.5$ or with $I < 3.2$ and $V < 6$; 36 of the stars are not included in Gaia EDR3.
The bright target stars are exposed through a small 12.5 neutral-density spot, allowing for unsaturated images of the target stars to be calibrated directly against a dense set of stars with accurate Gaia astrometry more than 12 magnitudes fainter on the same image frame.  
Combining multiple UBAD observations for each star with Hipparcos-2, the median position error for UBAD stars is 1.9 mas in both right ascension and declination at the catalog epoch of 2017, with 90\% of the stars having errors less than 2.6 mas; systematic errors are estimated at 1 -- 3 mas.  For those stars that include Hipparcos-2 in their solution, the median error in proper motion is 0.045 mas~year$^{-1}$ in right ascension and 0.049 mas~year$^{-1}$ in declination, with 90\% of stars having errors less than 0.1 mas~year$^{-1}$.  

Coupled with a faint extension that is not included in the UBAD catalog, UBAD observations overlap Gaia in the magnitude range $2 \lesssim G \lesssim 6$, extending from nearly the brightest stars to the non-saturated regime for Gaia.  It is not possible to calibrate any optical distortion under the neutral-density spot relative to the rest of the field of view, thus we can not constrain the possibility of an overall systematic error in Gaia in this magnitude range.  However, no magnitude-dependent systematic offset between Gaia and UBAD is evident at the milliarcsec level;  as the comparison extends into the non-saturated regime for Gaia, an overall systematic error in Gaia is unlikely.  After correction for the median offset between UBAD and Gaia,
UBAD positions are consistent with Gaia to approximately 2~mas, providing external validation of Gaia's bright-star astrometry to that level.

Gaia fundamentally changes the nature of ground-based astrometry.  By providing a dense set of calibrating stars with sub-milliarcsecond positions, it changes the field from one dominated by errors in reference catalogs to one dominated by instrumentation and atmospheric limitations. There remains a role for ground-based astrometry with small telescopes in the Gaia era.  A telescope such as the 61-inch, which has been producing sub-milliarcsecond differential astrometry for decades, can now produce milliarcsecond or better absolute astrometry, with the potential for significant impacts in the study of solar system bodies, resident space objects, and similar fields with astrometric requirements beyond what Gaia provides.

\begin{acknowledgements}
We gratefully acknowledge the work of the outstanding engineering group at USNO, Flagstaff Station (some now retired), including Mike Divittorio, Fred Harris, Mike Schultheis, Al Rhodes, and Andrew Cenko.  We also thank Bob Zavala for helpful comments on the text.

This work has made use of data from the European Space Agency (ESA) mission
{\it Gaia} (\url{https://www.cosmos.esa.int/gaia}), processed by the {\it Gaia}
Data Processing and Analysis Consortium (DPAC,
\url{https://www.cosmos.esa.int/web/gaia/dpac/consortium}). Funding for the DPAC
has been provided by national institutions, in particular the institutions
participating in the {\it Gaia} Multilateral Agreement.

The Pan-STARRS1 Surveys (PS1) and the PS1 public science archive have been made possible through contributions by the Institute for Astronomy, the University of Hawaii, the Pan-STARRS Project Office, the Max-Planck Society and its participating institutes, the Max Planck Institute for Astronomy, Heidelberg and the Max Planck Institute for Extraterrestrial Physics, Garching, The Johns Hopkins University, Durham University, the University of Edinburgh, the Queen's University Belfast, the Harvard-Smithsonian Center for Astrophysics, the Las Cumbres Observatory Global Telescope Network Incorporated, the National Central University of Taiwan, the Space Telescope Science Institute, the National Aeronautics and Space Administration under Grant No. NNX08AR22G issued through the Planetary Science Division of the NASA Science Mission Directorate, the National Science Foundation Grant No. AST-1238877, the University of Maryland, Eotvos Lorand University (ELTE), the Los Alamos National Laboratory, and the Gordon and Betty Moore Foundation.

This research has made use of the SIMBAD database,
operated at CDS, Strasbourg, France \citep{2000A&AS..143....9W}.
\end{acknowledgements}

\facility{USNO:61in (ND12)}

\software{astroalign 2.3.1 \citep{BEROIZ2020100384}, Astrometry.net 0.80 \citep{2010AJ....139.1782L}, Astropy 4.2 \citep{astropy:2013, astropy:2018}, Astroquery 0.4.1 \citep{2019AJ....157...98G}, PyERFA 1.7.1.1 \citep{https://doi.org/10.5281/zenodo.4279480}, Matplotlib 3.3.2 \citep{Hunter2007}, NumPy 1.19.4 \citep{2020NumPy-Array}, PALpy 1.8.1 \citep{2013ASPC..475..307J}, SciPy 1.5.3 \citep{2020SciPy-NMeth}, SExtractor 2.19.5 \citep{1996A&AS..117..393B}}

\bibliography{ubad}{}
\end{document}